\documentclass[amsmath,twocolumn,superscriptaddress,prx,aps]{revtex4-1} 
\usepackage{amsthm,amsfonts,graphicx,verbatim, color}
\usepackage{bm}
\usepackage[utf8]{inputenc}
\usepackage{ulem}
\usepackage{hyperref}
\usepackage{color}
\let\oldmarginpar\marginpar
\renewcommand\marginpar[1]{\-\oldmarginpar[\raggedleft\footnotesize #1]%
{\raggedright\footnotesize #1}}

\newcommand{\be}{\begin{equation}}
\newcommand{\ee}{\end{equation}}
\newcommand{\bea}{\begin{eqnarray}}
\newcommand{\eea}{\end{eqnarray}}

\renewcommand{\epsilon}{\varepsilon}

\hypersetup{
    colorlinks=true,
    linkcolor=blue,
    filecolor=blue,
    urlcolor=blue,
    citecolor=blue,
}

\newcommand{\addVR}[1]{{\color{blue}}}

\def\beq{\begin{equation}}
\def\eeq{\end{equation}}
\def\bea{\begin{eqnarray}}
\def\eea{\end{eqnarray}}

\begin{document}

\title{Anderson Localization and Swing Mobility Edge in Curved Spacetime}

\author{Shan-Zhong Li}
\affiliation{Guangdong Provincial Key Laboratory of Quantum Engineering and Quantum Materials, School of Physics and Telecommunication Engineering, South China Normal University, Guangzhou 510006, China}

\affiliation{Guangdong-Hong Kong Joint Laboratory of Quantum Matter, Frontier Research Institute for Physics, South China Normal University, Guangzhou 510006, China}

\author{Xue-Jia Yu}
\affiliation{International Center for Quantum Materials, School of Physics, Peking University, Beijing 100871, China}

\author{Shi-Liang Zhu}
\affiliation{Guangdong Provincial Key Laboratory of Quantum Engineering and Quantum Materials, School of Physics and Telecommunication Engineering, South China Normal University, Guangzhou 510006, China}

\affiliation{Guangdong-Hong Kong Joint Laboratory of Quantum Matter, Frontier Research Institute for Physics, South China Normal University, Guangzhou 510006, China}

\affiliation{Hefei National Laboratory, Hefei 230088, People's Republic of China}

\author{Zhi Li}
\email{lizphys@m.scnu.edu.cn}

\affiliation{Guangdong Provincial Key Laboratory of Quantum Engineering and Quantum Materials, School of Physics and Telecommunication Engineering, South China Normal University, Guangzhou 510006, China}

\affiliation{Guangdong-Hong Kong Joint Laboratory of Quantum Matter, Frontier Research Institute for Physics, South China Normal University, Guangzhou 510006, China}

\date{\today}
\begin{abstract}
We construct a quasiperiodic lattice model in curved spacetime to explore the crossover concerning both condensed matter and curved spacetime physics. We study the related Anderson localization and find that the model has a clear boundary of localized-extended phase separation, which leads to a swing mobility edge, i.e., the coexistence of localized, swing and sub-extended phases. The swing mobility edge, first reported here, features the phase-dependent eigenstate, that is, the eigenstate swing between the extended and localized state for differnt phase parameter of the quasiperiodic potential. Furthermore, A novel self-consistent segmentation method is developed to calculate the analytical expression of the critical point of phase separation, and the rich phase diagram is obtained by calculating the fractal dimension and scaling index in multifractal analysis.
\end{abstract}

\maketitle

\section{Introduction}
The past few decades have witnessed extensive studies on the Anderson localization~\cite{GFeher1959a,GFeher1959b,PWAnderson1958,NFMott1967,EAbrahams2010, PALee1985}, with many important results being achieved both theoretically~\cite{DVollhardt1980a,DVollhardt1980b,PALee1981,DVollhardt1982,SFishman1982,TGiamarchi1988,RGade1993,IEPsarobas2000,SESkipetrov2018,GLemarie2019,TJuntunen2019,FSuzuki2021,YSharabi2021} and experimentally~\cite{MCutler1967,SHikami1981,RLWeaver1990,RDalichaouch1991,FScheffold1999,AAChabanov2000,PPradhan2000,TSchwartz2007,LFallani2007,JBilly2008,ASPikovsky2008,GModugno2010,MSegev2013,WSchirmacher2018,BNagler2022,XCui2022}. To explain the disappearance of spin diffusion for low doping density~\cite{GFeher1959a,GFeher1959b}, P. W. Anderson proposed the famous theory of Anderson localization~\cite{PWAnderson1958}, which states that  the ergodic property of electrons in a system without interaction will be destroyed and the system will transform into a localized phase when the intensity of disorder exceeds the critical value, thus making the system turn from a metallic to an insulating phase. N. F. Mott further proposed the concept of mobility edge~\cite{NFMott1967}, which indicates that the localized and extended phases of the system can coexist under certain circumstances. Previous studies have suggested that one-dimensional (1D) and two-dimensional (2D) systems will exhibit localized behavior when uncorrelated disorder potential is introduced. In the three-dimensional (3D) case, however, the introduction of disorder will cause mobility edge, i.e., there appear both localized and extended states in the system~\cite{EAbrahams1979,PALee1985,FEvers2008}.

In addition to disordered systems, the 1D Aubry-Andr\'e-Harper (AAH) model, as a typical quasiperiodic system, has been eye-catching as one of the simplest systems to demonstrate the localized-extended phase transition~\cite{PGHarper1955,SAubry1980}. The quasiperiodic potential is incommensurate with the lattice space, which can be regarded as a limbo system between disorder and order. Due to the self-duality of AAH model, the system is characterized by an extended phase (localized phase) when the quasiperiodic potential is less than (greater than) the critical value, and the corresponding eigenstates are all in extended (localized) states~\cite{SAubry1980}. Therefore, compared with low-dimensional random disordered systems, quasi-periodic systems can exhibit the localized and extended phase transitions more efficiently. The AAH model is also valuable in studying topological phases in quasicrystals for the reason that it can be mapped to the 2D integer quantum Hall effect by a continuous U(1) gauge transformation~\cite{FMei2012,YEKraus2012a,YEKraus2012b,LJLang2012a,LJLang2012b,XCai2013,SGaneshan2013,HJiang2019,QBZeng2020,GQZhang2021,LZTang2021,LZTang2022,YPWu2022}. Apart from the standard AAH model, the studies on novel quasiperiodic systems have become a hot topic, where the mobility edge can be obtained by introducing a long-range correlation~\cite{XDeng2019,NRoy2021} or reconstructing the quasiperiodic potential~\cite{SDSarma1988,SGaneshan2015,XLi2017,HYao2019,YWang2020,DDwiputra2022,YCZhang2022}. Furthermore, many-body localization can be studied by exerting interaction~\cite{SIyer2013,VMastropietro2015,SZhang2018,YYoo2020,DWZhang2020}. So far, the quasiperiodic system has been realized experimentally in various platforms~\cite{GRoati2008,HPLuschen2018,MSchreiber2015,PBordia2017}.

On the other hand, in 1981, G.E. Unruh proposed a sonic horizon, which was the first attempt to simulate a black hole horizon and the relevant CST physics in the laboratories~\cite{WGUnruh1981,NMario2002}. The seminal work provides an effective way in exploring black holes, CST physics and general relativity. After 40 years of intensive efforts, long-lived black hole horizon and CST have been successfully simulated in various tabletop experiments, such as water flume~\cite{WGUnruh1981,RSchutzhold2002,SWeinfurtner2011,LPEuve2016,LPEuve2020}, Bose-Einstein condensates~\cite{LJGaray2000,OLahav2010,JSteinhauer2014,JSteinhauer2016,JRMunozdeNova2019},  exciton-polaritons~\cite{HSNguyen2015}, and nonlinear optics~\cite{TGPhilbin2008,IISmolyaninov2010,FBelgiorno2010,CCiret2016,YHWang2017,JDrori2019}. These milestone achievements provide a solid platform for studying the CST physics, which has deepened our understanding of the nature of gravity, e.g., helping us to reveal the relation between (1+1)D Jackiw-Teitelboim gravity and Sachdev-Ye-Kitaev model~\cite{SSachdev1993,AKitaev2015,AKitaev2018,JMaldacena2016a,JMaldacena2016b}. So far, despite the successful simulation of Hawking radiation, lots of effort on the verification of Unruh effect~\cite{WGUnruh1981,NMario2002,RSchutzhold2002,SWeinfurtner2011,LPEuve2016,LPEuve2020,LJGaray2000,OLahav2010,JSteinhauer2014,JSteinhauer2016,JRMunozdeNova2019,HSNguyen2015,TGPhilbin2008,IISmolyaninov2010,FBelgiorno2010,CCiret2016,YHWang2017,JDrori2019,SSachdev1993,AKitaev2015,AKitaev2018,JMaldacena2016a,JMaldacena2016b,SSchlicht2004,LCBCrispino2008,EMartinMartinez2011} and other phenomena related to CST physics~\cite{LCBCrispino2008}, just a few researches have been focused on the condensed matter properties in CST lattice system~\cite{AWeststrom2017,YKedem2020,CMorice2021,GEVolovik2021,CMorice2022,DSabsovich2022,OBoada2011,BMula2021,VFaraoni2015,JRodrguezLaguna2017,FZhong2018,CSheng2021,LMertens2022}.

Inspired by the aforementioned achievements in artificial CST systems, we construct a generalized 1D quasiperiodic lattice model in CST to explore the crossover concerning both condensed matter and CST physics. We find that condensed matter lattice systems with Anderson phase transition will exhibit phase separation in CST. Besides, a novel ``segmentation'' method is developed to obtain the analytical expression of the critical position of the phase separation. Based upon the method, we reveal that the swing mobility edge emerges in the system, i.e., localized, swing and sub-extended phase can coexist in the system. Different from conventional quasiperiodic systems, CST-AAH model exhibits different localization characteristics for different phase angles, i.e., corresponding eigenstates show swing behavior, which means eigenstates will exhibit extended or localized features for differet phase angles. This work provides a CST-version of Anderson localization and mobility edge theory.

\begin{figure}[htbp]
	\centering
	\includegraphics[width=8.2cm]{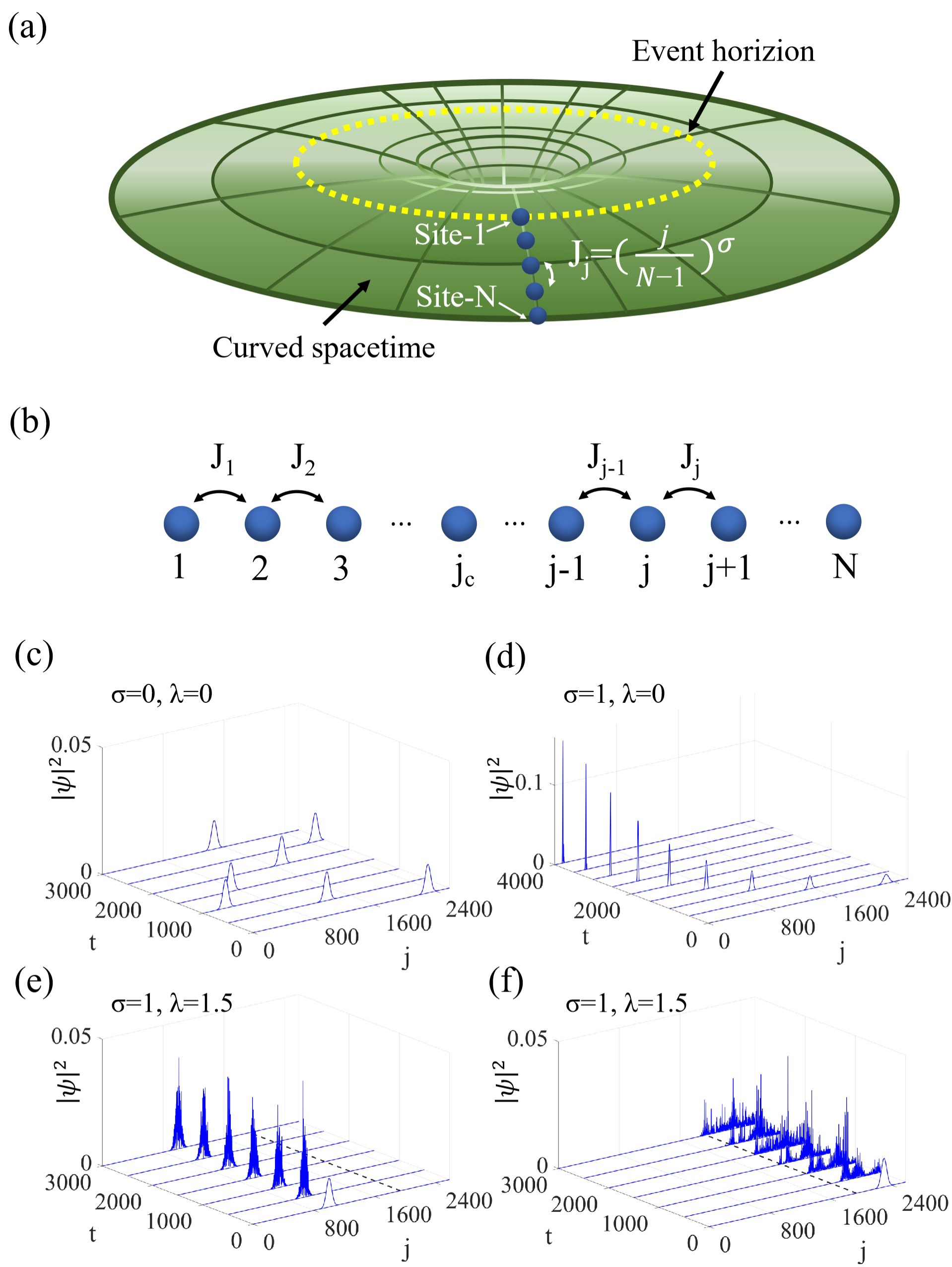}
	\caption{(Color online). (a) Schematic diagram of the CST-AAH model in the vicinity of a black hole, where the nearest neighboring hopping is power-law position-dependent. (b) AAH chain in CST shows phase separation, while the whole chain is divided into the localized ($j<j_{c}$) and extended ($j>j_{c}$) regions. (c)-(f) The evolution of a test wavepacket in flat space ($\sigma=0$) and in CST ($\sigma=1$). The first (second) row corresponds to the case where the quasiperiodic potential is absent (present). The centers of the initial states are at $j_0=1000$ and $j_0=2300$, respectively. The system size $N=2584$ and the corresponding critical site (dashed line) of phase separation $j_{c}=1937$.} \label{Sch}
\end{figure}

\section{Model}
Based on the CST lattice model~\cite{AWeststrom2017,CMorice2021,BMula2021,CSheng2021,JRodrguezLaguna2017}, we construct a CST version of quasiperiodic system as shown in Fig.~\ref{Sch}(a)(b). The corresponding Hamiltonian reads (see Appendix~\ref{APA} for details)
\begin{equation}\label{HAMI}
H=\sum_{j=1}^{N-1} J_{j}(\hat{c}^{\dag}_{j}\hat{c}_{j+1}+\mathrm{H.c.})+\sum_{j=1}^{N}V_{j}\hat{c}^{\dag}_{j}\hat{c}_{i},
\end{equation}
where $\hat{c}^{\dagger}_{j}\ (\hat{c}_{j})$ is the fermionic creation (annihilation) operator at the $j$th site. The on-site potential $V_{j}=\lambda \cos(2\pi \phi  j+\theta )$, where $\lambda$ denotes the strength of the incommensurate potential, $\phi$ is an irrational number, and $\theta\in\left[0 ,2\pi \right]$ is a phase angle~\cite{DWZhang2018}. For CST, the nearest neighboring hopping strength $J_{j}=J(\frac{j}{N-1})^{\sigma}$, which depends on the site index $j$ and the parameter $\sigma$ indicating warping degree of spacetime~\cite{CMorice2021}, i.e., the larger the value of $\sigma$, the greater the warping of spacetime. When $\sigma=0$, the spacetime returns to a flat one. When $N$ is large enough and $\sigma>1$, $J_{1}\rightarrow0$, system presents a horizon from the first site to the critical point, where information cannot pass through. Therefore, to simulate the behavior outside the event horizon, we use open boundary conditions. Without loss of generality, hereafter we take $J=1$ as the unit of energy, and select $\phi=(\sqrt{5}-1)/2$ as the typical irrational number. 

The standard AAH model (without spacetime warping for the condition of $\sigma=0$ and thus $J_j=J$) exhibits a phase transition at $\lambda_c =2J$~\cite{SAubry1980}, i.e., the system is of extended (localized) phase for $\lambda<\lambda_c$ ($>\lambda_c$). Properties induced by CST emerge when $\sigma\geq1$, where the corresponding hoppings gradually increase from $0$ to $1$ with the lattice index growing from small to large~\cite{CMorice2021,YKedem2020,CMorice2022}. As is known, nothing, even as minuscule as photons, can be spared from being pulled in the vicinity of a super gravitational source. The closer the little thing is to the gravitational monster, the greater the influence it will feel. Therefore, it is reasonable to assume that for the near-end of AAH chain to the event horizon, hopping becomes difficult due to the extreme attraction, while the rear-end of the chain, farther away from the gravitational pull, exhibits normal flat-spacetime hopping.

Therefore, the CST-AAH chain [Eq.~\eqref{HAMI}] can well reflect the lattice-gravity correspondence and the relevant dynamical properties in the vicinity of an event horizon.  The model proves an ideal simulator to reveal the event horizon dynamics of free particles in (1+1)D anti-de Sitter space, where the particles slow down exponentially as they move towards the event horizon, and vice versa~\cite{NMario2002}.

\section{Phase Separation}
Although the analytical expression of mobility edge for the AAH model with constant hopping strength ($J_{j}=J$) can be obtained by Avila's global theorem~\cite{AAvila2015}, the method does not work for a general $J_{j}$ in CST. Here, we propose a brand new method of segmentation to explore the localization properties of CST-AAH model. By considering a chain composed of two subchains, four subchains and N-1 subchians, we eventually approximate to the extreme case: CST-AAH chain (see Appendix~\ref{APB} for details). Based on the known critical point of the phase transition in the flat standard AAH model, we analytically solve the CST-AAH model. The results show that the phase separation will occur, which divides the whole AAH chain into two parts (the localized and the extended) with a clear boundary in between. As shown in Appendix~\ref{C1}, the analytical expression of phase separation's critical site $j_c$ can be obtained as
\begin{equation}
\label{jc}
j_{c}=\left \lfloor \left(\frac{\lambda}{2J}\right)^{\frac{1}{\sigma}}(N-1) \right \rfloor.
\end{equation}
Here $\lfloor...\rfloor$ denotes floor function, which is defined to round down the number inside the function to an integer. When $\sigma=0$, the system reduces to the case of flat spacetime, and the expression Eq.~\eqref{jc} becomes $j_{c}=\left\lfloor \left(\frac{\lambda}{2J}\right)^{\infty}(N-1) \right\rfloor$. One can see that as $\lambda$ increases, the whole system will exhibit extended phase ($0<\frac{\lambda}{2J}<1$) first and then localized ($\frac{\lambda}{2J}>1$) phase. Before and after the threshold value $\lambda=2J$, the critical site $j_c=0$ and $j_c=\infty$, which indicates no coexistent localized and extended phases. That is to say, the system can only be of a pure extended or a pure localized state, which is consistent with what we knew previously on standard AAH chain (Appendix~\ref{C1}). Considering the opposite extreme case, when spacetime is infinitely curved ($\sigma=\infty$), one can get $j_{c}=\left\lfloor (N-1) \right\rfloor$. One can learn from the expression that the boundary of phase separation is always at the rightmost end of the chain, and then the whole chain of the system exhibits the localized phase, which is quite in line with our knowledge: infinitely curved spacetime means everything frozen.

Wavepacket dynamics is a very effective way to reflect the CST properties~\cite{AWeststrom2017,YKedem2020,CMorice2021}. One can consider a general Gaussian function as the initial state, i.e.,
\begin{equation}
    \psi(j,t=0)=\frac{1}{\sqrt[4]{\pi}\sqrt{w}}e^{-\frac{1}{2}(\frac{j-j_{0}}{w})^2}e^{ip_{0}j},
\end{equation}
where the width of the wavepacket $w=50$, and the initial momentum $p_0=-\pi/2$ in the numerical calculation. The results are plotted in Fig.~\ref{Sch}(c)-(f), and calculation details are contained in Appendix~\ref{C1}. As a comparison, we show the cases with no quasiperiodic potential first, i.e., $\lambda=0$. The results show that the wavepacket in the flat spacetime [Fig.~\ref{Sch}(c)] is extended over the entire chain, while that in CST behaves more like an object on the verge of the black hole horizon, featuring continuous deceleration and localization [Fig.~\ref{Sch}(d)]. The presence of quasiperiodic potential in the system can give rise to a novel phenomenon of phase separation, with a clear boundary $j_c$ existing between the localized and the extended regions [dashed line in Fig.~\ref{Sch}(e)(f)]. The wave function exhibits the localized (extended) characteristics if it is initially placed in the localized (extended) region.

\section{Mobility Edge}
In order to explore the phase seperation and corresponding localization properties of CST-AAH chain, we calculate the fractal dimension and scaling indices, both of which are core observables in the investigation of the localization and mobility edge.

\begin{figure}[thbp]
	\centering
	\includegraphics[width=8.2cm]{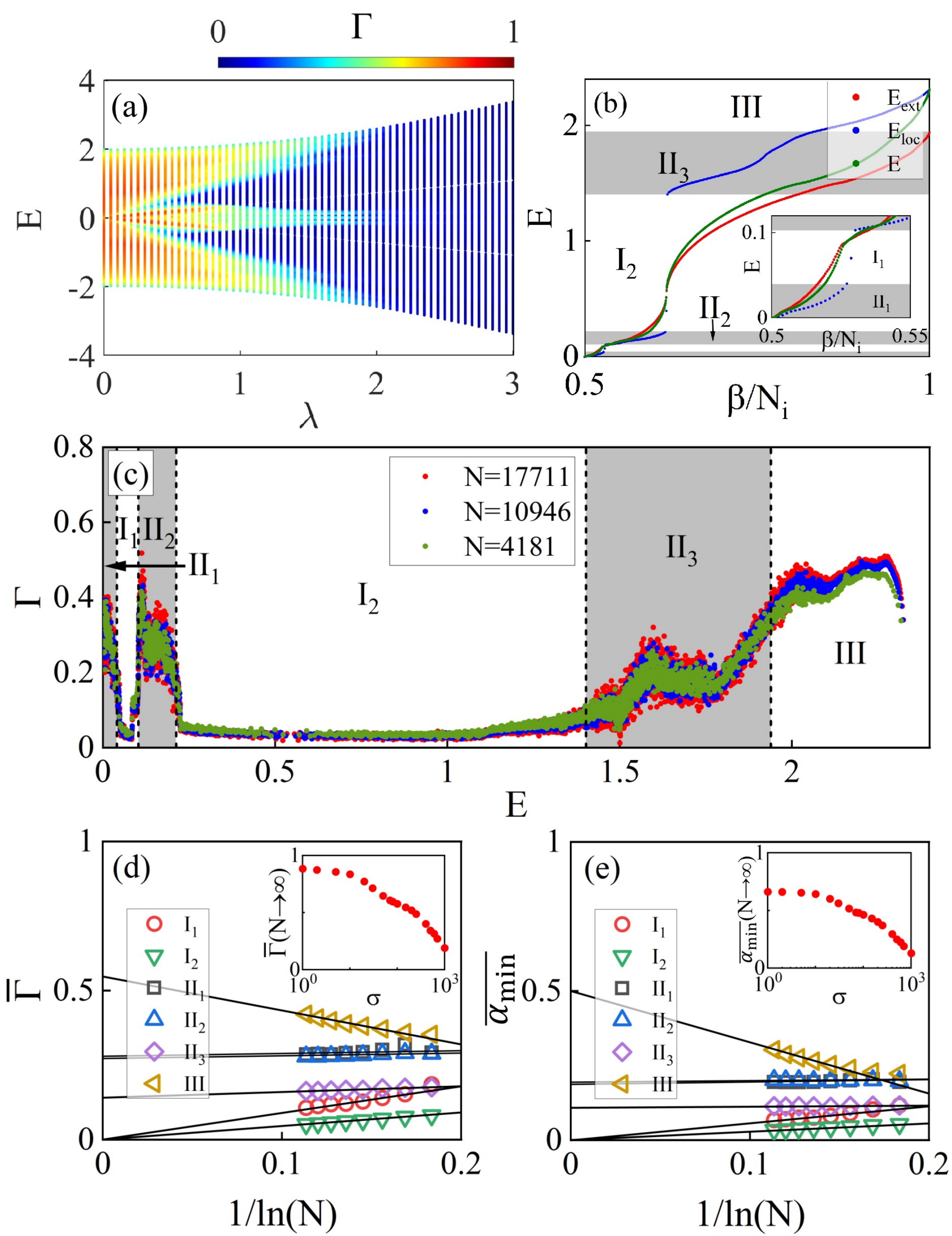}
	\caption{(Color online). (a) Fractal dimension $\Gamma$ of all wave functions for CST-AAH model with $\sigma=1$. (b) The distribution of eigenvalues versus level index $\beta$ for extended and localized subchains, respectively. $N_{i}$ is the chain or subchain size, where $i=ext,~loc,~total$, corresponds to the extended subchain  size $N-j_c$, the localized subchain  size $j_{c}$, and the total chain size $N$, respectively. In the gray regions, the eigenvalues of the extended and localized subchains overlap, which is the evidence of the emergence of the swing phase. (c) The fractal dimension $\Gamma$ of eigenstates for different eigenvalues $E$ with sizes $N=4181$ (green), $N=10946$ (blue) and $N=17711$ (red). The other parameters $\sigma=1$, $\lambda=1.5$. The gray regions show swing phase. The scaling properties of $\overline{\Gamma}$ (d) and $\overline{\alpha_{min}}$ (e) as a function of $1/\ln(N)$ for different regions are provided. Insets: The effect of CST parameter $\sigma$ on $\overline{\Gamma}(N\rightarrow\infty)$ and $\overline{\alpha_{min}}(N\rightarrow\infty)$. The system size $N=2584$ in (a)(b), and 100 times quasiperiodic averages have been performed on $\theta$ for all plots.} \label{FD}
\end{figure}

Firstly, we calculate the fractal dimension defined as
\begin{equation}
\Gamma(\beta)=-\lim_{N\rightarrow \infty}\frac{\ln \xi(\beta)}{\ln N},
\end{equation}
where $\xi(\beta)=\sum_{j=1}^{N}\left | \psi_{j}(\beta)\right|^4 $ denotes the inverse participation ratio (IPR), and $\beta$ is the energy level index of the particle eigenstate. The fractal dimension $\Gamma\rightarrow0$ ($\rightarrow1$) corresponds to the localized (extended) state, while $\Gamma\in(0,1)$ to multifractal state~\cite{YWang2020,HYao2019,DDwiputra2022,YWang2022a,YWang2022b}. Previous studies suggested two possible ways to induce mobility edge in the AAH model, i.e., by introducing a long-range hopping or energy dependent quasiperiodic potential~\cite{XDeng2019,NRoy2021,SDSarma1988,SGaneshan2015,XLi2017,HYao2019,YWang2020,DDwiputra2022}. Here we show that warping the spacetime is another way to induce mobility edge [see Fig.~\ref{FD}(a)].

To better understand the generation mechanism of mobility edge shown in Fig.~\ref{FD}(a), we compute the subchain eigenvalues versus level index $\beta$ of localized and extended subchains, respectively. As shown in Fig.~\ref{FD}(b), the results reveal that there are three different phases: pure localized phase (region I$_{1}$ and I$_{2}$), swing phase (region II$_{1}$, II$_{2}$ and II$_{3}$), and pure extended phase (region III). On the one hand, the extended and localized phases correspond to pure subchain eigenvalues~\cite{XLin2022}, i.e., $E_{ext}$ and $E_{loc}$, corresponding Hamiltonians read
\begin{equation}
\begin{aligned}
&H_{ext}=\sum_{j=1}^{j_{c}-1} J_{j}(\hat{c}^{\dag}_{j}\hat{c}_{j+1}+\mathrm{H.c.})+\sum_{j=1}^{j_{c}}V_{j}\hat{c}^{\dag}_{j}\hat{c}_{j},\\
&H_{loc}=\sum_{j=1}^{N-j_{c}}J_{j_{c}-1+j}(\hat{c}^{\dag}_{j}\hat{c}_{j+1}+\mathrm{H.c.})+\sum_{j=1}^{N-j_{c}+1}V_{j_{c}-1+j}\hat{c}^{\dag}_{j}\hat{c}_{j},
\end{aligned}
\end{equation}
where the total number of localized (extended) subchain is $j_{c}$ ($N-j_{c}$) and the value range of corresponding hopping strength is from $J_{1}$ to $J_{j_{c}-1}$ ($J_{j_{c}}$ to $1$). On the other hand, the swing phase features the superposition of eigenvalues of two different subchains, i.e., the coexistance region of $E_{ext}$ and $E_{loc}$ [the gray regions of Fig.~\ref{FD}(b)]. 

Furthermore, one can distinguish different phases by the behavior of the fractal dimension versus system size. To this end, we perform the scaling analysis and plot the results of different $N$ in Fig.~\ref{FD}(c). One can find that the fractal dimension $\Gamma$ in region III (region I$_{1}$ and I$_{2}$) increases (decreases) with the increasing system size $N$, which exhibits the properties of the extended (localized) phase. However, $\Gamma$ is independent of the system size in regions II$_{1}$, II$_{2}$ and II$_{3}$. This phenomenon arises from the counteracting scaling behaviors between the extended and localized eigenstates within this domain, thereby leading to the size-independent nature of $\Gamma$, which is an evidence of the swing phase. Both the scaling behavior of fractal dimension and the overlap of subchain eigenvalues agree well with each other, and thus corroborate the emergence of the swing phases. Secondly, we calculate the scaling index in multifractal analysis to further explore different phases~\cite{HHiramoto1989,HGrussbach1995,SSchiffer2021, JWang2016}. The probability of a particle occupied in site $j$ is represented by the modulus square of the wave function $\mathbb{P}_{j}=\left| \psi_{j} \right|^2$, which satisfies the normalization condition $\sum_{j}\left| \psi_{j} \right|^2=1$. The scaling index of multifractal analysis $\alpha_{j}$ is defined by the probability measure $\mathbb{P}_{j}$ as
\begin{equation}
\mathbb{P}_{j}=N^{-\alpha_{j}}.
\end{equation}
Since the occupation probability on all sites is $\mathbb{P}_j =1/N$ for a completely extended wave function, the corresponding scaling index $\alpha_j=1$. For a localized wave function, the occupation probability is non-zero at just a few sites, therefore $\alpha\rightarrow 0$ for such occupied sites and $\alpha \rightarrow \infty$ for the other sites. For a multifractal wave function, the scaling index $\alpha$ is distributed in a finite interval $\left[\alpha_{min}, \alpha_{max} \right]$. Thus, by considering the thermodynamic limit $N \rightarrow \infty$, one can characterize the localization properties of a wave function by $\alpha_{min}$. To be specific, for $N \rightarrow \infty$, $\alpha_{min}=1~(0)$ indicates the extended (localized) states, whereas $0<\alpha_{min}<1$ corresponds to the multifractal state.

To better demonstrate the properties of the wave functions in different regions, a routine approach is to calculate the mean values of $\Gamma$ and $\alpha_{min}$ in different regions, which are defined as
\begin{equation}
\overline{\Gamma}=\frac{1}{\eta_{R}}\sum_{R}\Gamma, \ \ \ \ \
\overline{\alpha_{min}}=\frac{1}{\eta_{R}}\sum_{R}\alpha_{min},
\end{equation}
where $\eta_{R}$ denotes the total number of eigenstates in the region denoted as $R=$I$_{1},~$I$_{2},~$II$_{1},~$II$_{2},~$II$_{3}, ~$III. One can calculate the corresponding fractal dimensions and scaling indices with different region sizes and extrapolate the data to get $\bar{\Gamma}$ and $\bar{\alpha}$ under the thermodynamic limit~\cite{YWang2022a,YWang2022b,JWang2016}. The corresponding results are plotted in Fig.~\ref{FD}(d)(e). Under this condition, the values of $\overline{\Gamma}$ and $\overline{\alpha_{min}}$ in the localized region decrease with the growing lattice size and finally approach zero. However, the values of $\overline{\Gamma}$ and $\overline{\alpha_{min}}$ in the extended regions increase with the increasing lattice size until they approach a fixed value. Note that, for the AAH model in flat spacetime, both $\overline{\Gamma}$ and $\overline{\alpha_{min}}$ in the extended region will eventually be close to $1$. The mechanism behind the above interesting phenomenon lies in the phase separation featured by AAH chain in CST. Though the wave function of the extended state can experience all states of the extended subchain, it is all the way prohibited from entering the localized subchain. The insets of Fig.~\ref{FD}(d)(e) exhibit the $\overline{\Gamma}$ and $\overline{\alpha_{min}}$ of all eigenstates versus the CST parameter $\sigma$ at $\lambda=0$, and the results support that the larger the degree of spacetime warping, the smaller the value of $\overline{\Gamma}(N\rightarrow\infty)$ and $\overline{\alpha_{min}}(N\rightarrow\infty)$.

\begin{figure}[htbp]
\centering
\includegraphics[width=8.5cm]{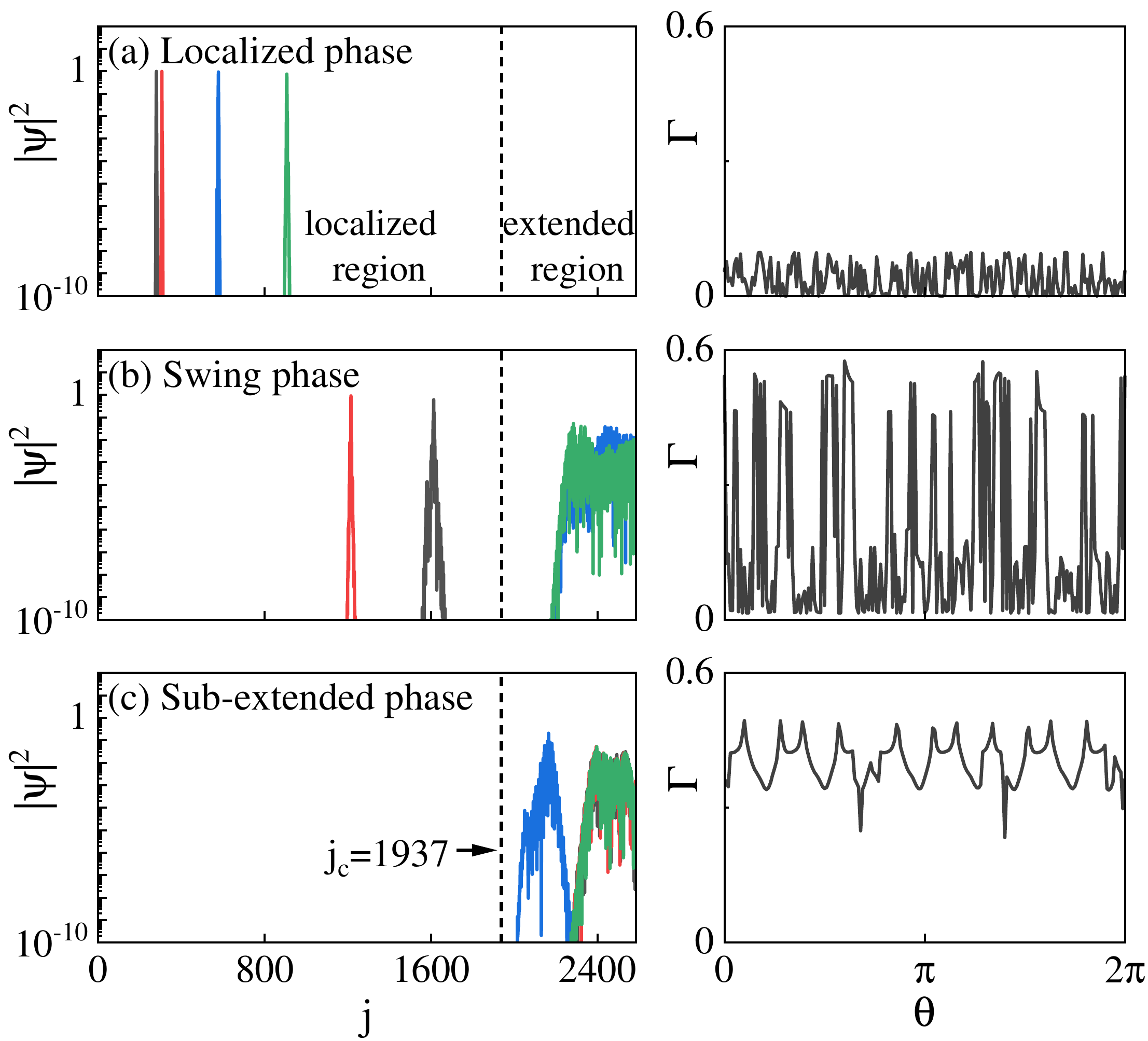}
\caption{(Color online). The probability distribution of eigenstates corresponding to the eigenvalues $E=1.00124$ (a), $E=1.63462$ (b), and $E=2.13195$ (c) with $\sigma=1$ and $\lambda=1.5$ for $\theta=5.1191$ (black), $3.9732$ (blue), $1.7499$ (green) and $0.7979$ (red). Dashed line shows the critical position of phase separation. The right-hand column shows the fractal dimension $\Gamma$ of the corresponding eigenstate as a function of $\theta$. The system size $N=2584$ and then $j_{c}=1937$.}\label{eig}
\end{figure}

\begin{figure*}[htp]
	\centering
	\includegraphics[width=15.5cm]{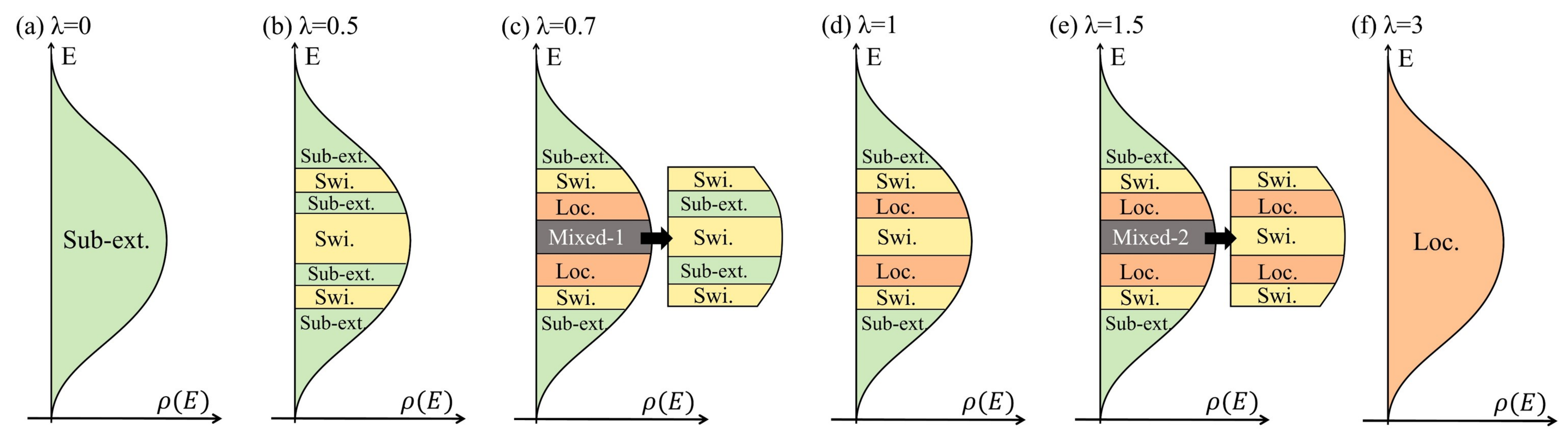}
	\caption{(Color online). The full phase diagram of AAH model in CST ($\sigma=1$), where Sub-ext., Swi. and Loc. are the abbreviations of Sub-extended, Swing and Localized phases, respectively. For clarity, the central regions in (c) and (e) are magnified by the insets. The corresponding detailed analysis is provided in Fig.~\ref{S8} and Fig.~\ref{SI} of Appendix~\ref{APF}.}\label{PD}
\end{figure*}

Furthermore, the distribution of the eigenstate wave functions can soundly reveal the effects of the localized-extended phase separations. Fig.~\ref{eig}(a)-(c) exhibit the distribution of eigenstate wave functions of three typical phases, and we randomly select four phase angle $\theta$ in the calculation. The results reveal that the wave functions of localized phase [Fig.~\ref{eig}(a)] remain localized regardless of the value of $\theta$, and the wave functions are all localized in the subchain of $j<j_c$. On the other hand, while the wave functions of the sub-extended phase [Fig.~\ref{eig}(c)] still display extended behaviors for different $\theta$, the extended state is confined to the region of extended subchains ($j>j_c$). For wave functions of the swing phase [Fig.~\ref{eig}(b)], localization characteristics of the system depend on the value of parameter $\theta$. In other words, different values of $\theta$ may produce either extended or localized states, which represents a whole new style of swing phase that has not been reported. We exhibit the variation of the fractal dimension $\Gamma$ of corresponding eigenstates with respect to the phase angle $\theta$ in the right-hand column to better illustrate this phenomenon. It can be seen clearly that the localization properties of the sub-extended and localized states do not change significantly versus $\theta$, while the fractal dimension of the swing state switches between the two. After averaging of different $\theta$, although the value is between the localized state and the sub-extended state, which is similar to the multifractal case, the system is actually in a brand new ``swing phase''. In other words, the numerical results are similar to the multifractal state, but it is actually an average behavior of sub-extended and localized states. Therefore, we call it the ``swing'' state. 

Finally, the full phase diagram of CST-AAH model is obtained as shown in Fig.~\ref{PD}. One can see that with the increase of the quasiperiodic potential parameter $\lambda$, the system experiences four intermediate phases from the extended to the final localized state [Fig.~\ref{PD}(b)-(d)]. Stepwise analyses of Fig.~\ref{PD} are given as follows. When $\lambda=0$, the whole system resides in the sub-extended phase [Fig.~\ref{PD}(a)], while the localized properties become increasingly salient as $\lambda$ grows larger. First, there appears a multilayered structure composed of the sub-extended and swing phases [Fig.~\ref{PD}(b)]. Then $\lambda$ continues to grow, leading to a much richer phase diagram that contains the sub-extended, swing and localized phases [Fig.~\ref{PD}(c)]. With the ever-increasing $\lambda$, the localized properties gradually gain the upper hand [Fig.~\ref{FD}(d)], occupying an overwhelming majority of regions in the multilayered structure [Fig.~\ref{PD}(e)]. Eventually, when $\lambda$ exceeds the critical value, the entire AAH chain will become localized [Fig.~\ref{PD}(f)].

\section{Conclusion}
In summary, we have constructed a CST-AAH model to explore the properties of condensed matter in CST. We found a phase separation phenomenon of CST-AAH model with a clear boundary, where the entire AAH chain can be regarded as a combination of the localized and the extended subchains. By applying the segmentation method, the analytical expression of phase separation of the critical position was obtained. Furthermore, we found that the phase separation gives rise to a swing mobility edge, i.e., the localized, swing and sub-extended phases coexist in the system, where the eigenstates of the swing phase may be either sub-extended or localized for different phase parameter $\theta$. In CST, it is impossible to be fully expanded even for the wave function originally in the extended subchain, hence we call the state ``sub-extended state'' . Our work is devoted to constructing the CST-version of Anderson localization and mobility edge theory, helping foster a crossover research concerning both condensed matter and CST physics. Nowadays, the ever advancing experimental techniques have enabled black hole horizons and CST to be simulated in various artificial systems~\cite{DWZhang2018,CBarcelo2019,CViermann2022}. Thus, it is promising that the phenomenon predicted here will be realized in experiments in the near future.

\section*{Acknowledgements}
We thank Yu-Cheng Wang, Dan-Wei Zhang, Li-Jun Lang, Lei Yin and Xi-Dan Hu for helpful discussions and constructive suggestions. This work was supported by the National Key Research and Development Program of China (Grant  No. 2022YFA1405300), the National Natural Science Foundation of China (Grant  No. 12074180), the Innovation Program for Quantum Science and Technology (Grant no. 2021ZD0301705), and the Guangdong Basic and Applied Basic Research Foundation (Grants No.2021A1515012350).

\appendix

\section{The construction of a quasiperiodic model in curved spacetime}\label{APA}
As suggested by refs.~\cite{CMorice2021,CMorice2022}, to establish a connection between continuum field theory and the condensed matter, one can start from 1D Jackiw-Teitelboim gravitation gauge with dilaton scalar field to obtain a 1D CST lattice model with position-dependent hopping strength at last, where the corresponding Hamiltonian reads
\begin{equation}\label{SE1}
H=\sum_{j=1}^{N-1} J_{j}(\hat{c}^{\dag}_{j}\hat{c}_{j+1}+\mathrm{H.c.}),
\end{equation}
where $\hat{c}^{\dag}_{j}$ and $\hat{c}_{j}$ correspond to the creation and annihilation operators, respectively, and $J_{j}=J(\frac{j}{N-1})^{\sigma}$ denotes the hopping strength between site $j$ and site $j+1$. It can be seen that Eq.~\eqref{SE1} is the discrete version of a Hamiltonian for a Dirac fermion on curved $(1+1)$D spacetime with a static metric of the form
\begin{equation}\label{MR}
    ds^2=-J^2(x)dt^2+dx^2.
\end{equation}
For the appropriate coordinates $x$ and $t$, the spacetime can be described by the above Rindler metric, with a position-dependent speed of light. One can define 
\begin{equation}
d\tilde{x} =dx/J(x)
\end{equation}
to obtain the equivalent Minkowski metric
\begin{equation}
ds^{2} =J^{2}(x)(-dt^2+d\tilde{x}^2),
\end{equation}
which is conformally equivalent to the metric Eq.~\eqref{MR}. We let $J(x)=J(\frac{j}{N-1})^{\sigma}$. When $J(x)=0$, the local speed of light disappears and information cannot pass through from there, thus separating spacetime into two Rindler wedges. In this paper, the same spacing is used in the diagram in order to demonstrate phase separation. Conformal equivalence between two metrics suggests that conformal field theory techniques would describe the universal properties of low-energy eigenstates of Hamiltonian Eq.~\eqref{SE1}. By the equivalence principle, any casual horizon can be approximated by the Rindler metric in a small region of spacetime~\cite{OBoada2011,BMula2021,VFaraoni2015,JRodrguezLaguna2017,FZhong2018,CSheng2021,LMertens2022}, such as the spacetime structure close to a Schwarzschild black hole horizon~\cite{OBoada2011,JRodrguezLaguna2017,PCWDavies1975,DGBoulware1975}. 

Under the thermodynamic limit ($N\rightarrow\infty$), the hopping strength of the two nearest neighboring sites can be regarded as a constant, thus one can obtain an approximate localized band structure, i.e., $\varepsilon (j,k)\approx -2(j/N)^{\sigma}\cos{k}$. Therefore, the corresponding dispersion relation of the Hamiltonian Eq.~\eqref{SE1} at $k=\pm\pi/2$ has a Dirac cone shape, and its quasiparticle shows the Dirac fermionic property. The position-dependent group velocity of the quasiparticle is similar to that of the Dirac field in a 1D Jackiw-Teitelboim gravitational background~\cite{OBoada2011}. When $\sigma>0$, the quasiparticle's group velocity vanishes at the sites of $j\rightarrow0$, and the quasiparticle shows in its behavior the group velocity of the light cone approaching the event horizon of a black hole. One can capture this interesting phenomenon through the wavepacket evolution. We consider a general Gaussian initial state as follows,
\begin{equation}
    \psi(j,t=0)=\frac{1}{\sqrt[4]{\pi}\sqrt{w}}e^{-\frac{1}{2}(\frac{j-j_{0}}{w})^2}e^{ip_{0}j},
\end{equation}
where $w$ is the width of the wavepacket, $j_0$ is the position of the center of the wavepacket, and $p_0=-\pi/2$ is the initial momentum.

\begin{figure}[tbhp]
	\centering
	\includegraphics[width=8.5cm]{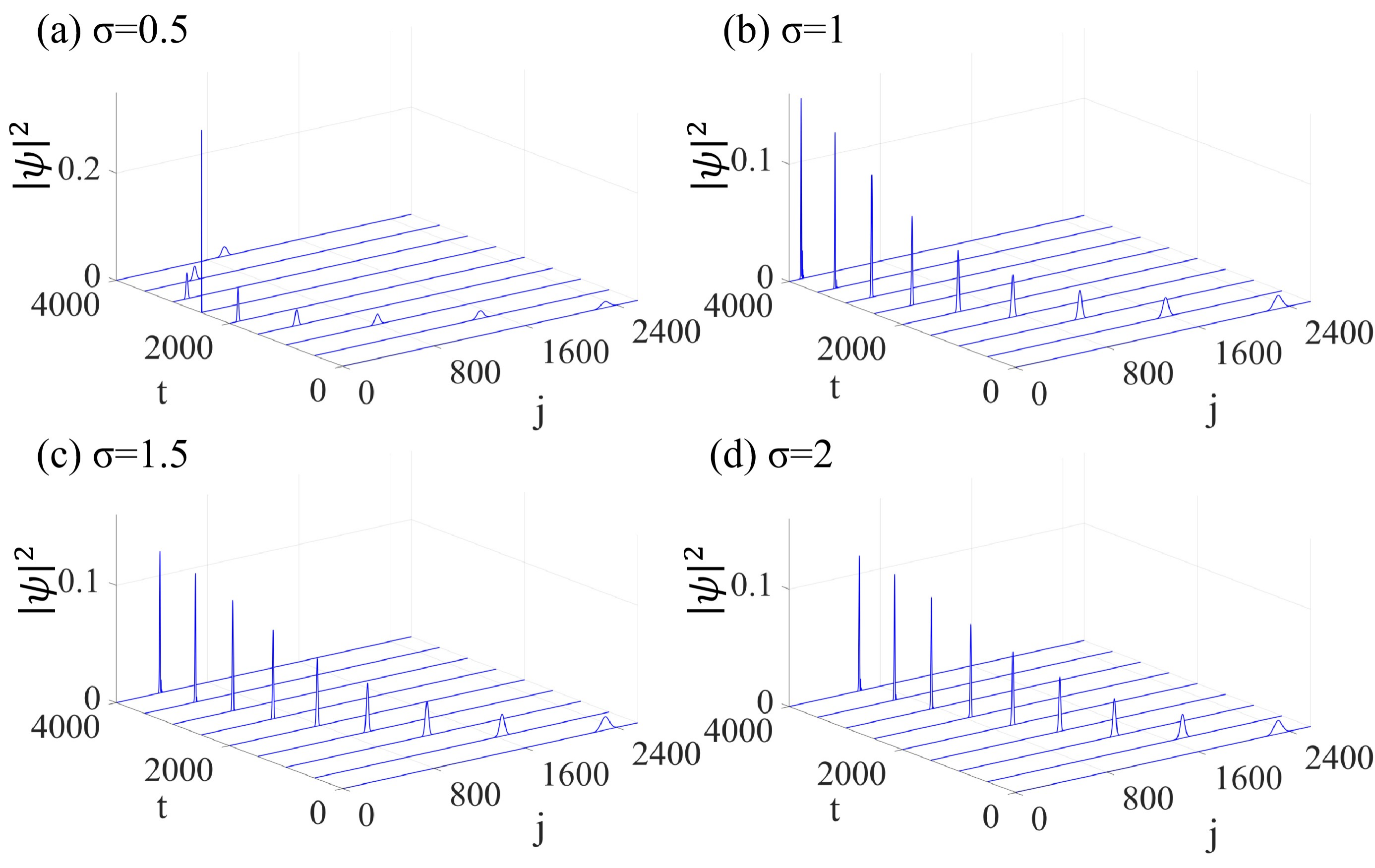}
	\caption{(color online). The wavepacket evolution of CST lattice model ($\lambda=0$) with $\sigma=0.5$ (a), $1$ (b), $1.5$ (c) and $2$ (d). The initial positions of the wavepacket are at $j_{0}=2300$. The system size $N=2584$.}\label{SMF1}
\end{figure}

We can see from Fig.~\ref{SMF1}(a) that, when $\sigma=0.5$, since the wavepacket bounces back after touching the boundary, it can not simulate the deceleration process of the wavepacket approaching a black hole horizon.

However, as shown in Fig.~\ref{SMF1}(b)-(d), the evolving wavepacket slows down at the sites $j\rightarrow0$ and resides near $j=0$. When $\sigma\geq1$, to be specific, the system can effectively simulate the dynamical properties of the wavepacket in the vicinty of the black hole horizon, i.e., the wavepacket becomes slower and more localized as it approaches the black hole horizon.

\begin{figure}[tbhp]
	\centering
	\includegraphics[width=8.5cm]{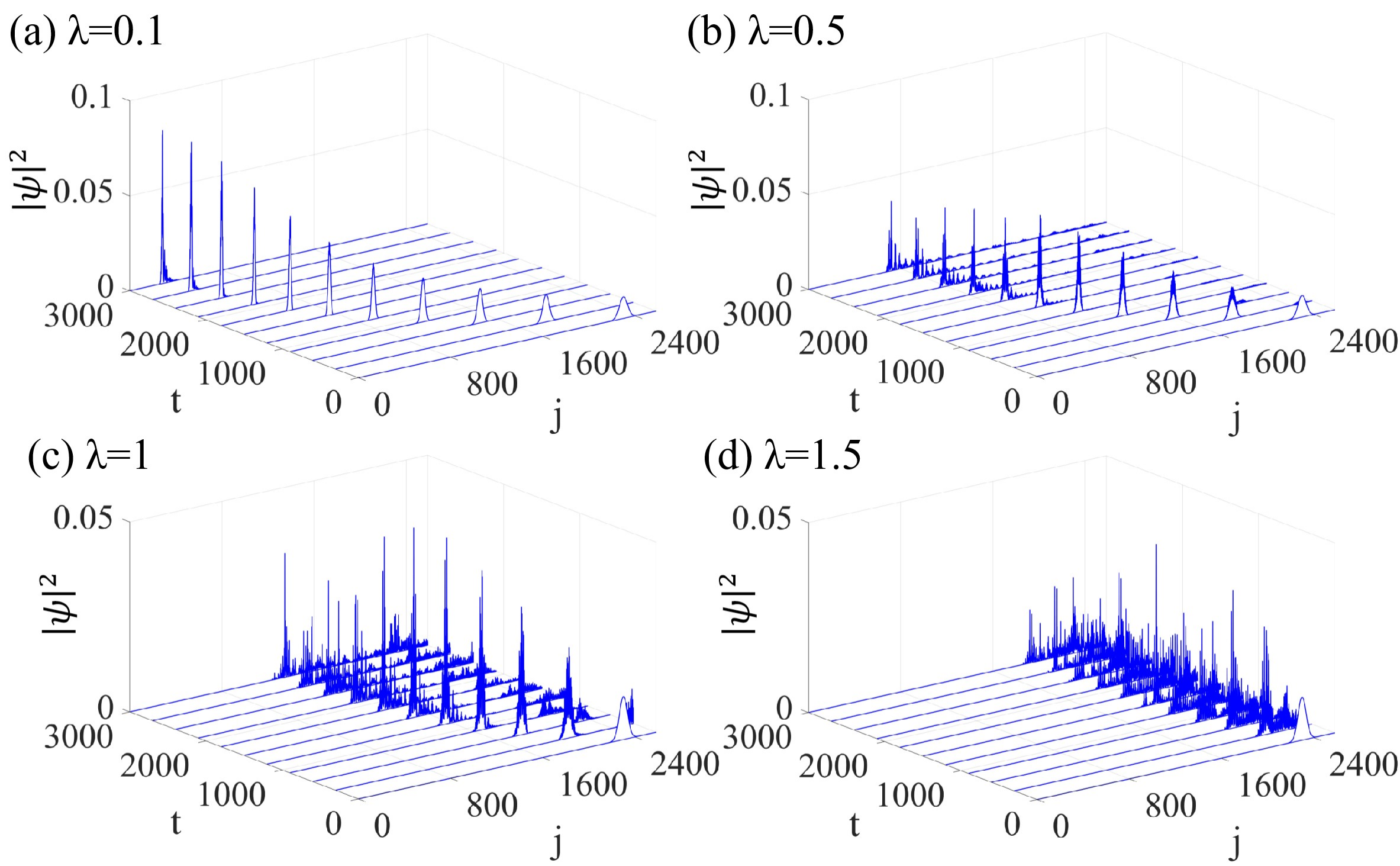}
	\caption{(color online). The wavepacket evolution of $\sigma=1$ CST-AAH model with (a) $\lambda=0.1$, (b) $\lambda=0.5$, (c) $\lambda=1$ and (d) $\lambda=1.5$. The initial positions of the wavepacket are at $j_{0}=2300$. The system size $N=2584$.}\label{SMF2}
\end{figure}

Based on the above analysis, we develop a quasiperiodic model in CST by applying quasiperiodic potential energy at each site, which can be used to explore the CST-version of Anderson localization theory. The corresponding Hamiltonian takes the form
\begin{equation}\label{SME3}
H=\sum_{j=1}^{N-1} J_{j}(\hat{c}^{\dag}_{j}\hat{c}_{j+1}+\mathrm{H.c.})+\sum_{j=1}^{N}V_{j}\hat{c}^{\dag}_{j}\hat{c}_{j},
\end{equation}
where $\hat{c}^{\dagger}_{j}\ (\hat{c}_{j})$ is the fermionic creation (annihilation) operator at the $j$th site. The on-site potential $V_{j}=\lambda \cos(2\pi \phi  j+\theta )$, where $\lambda$ denotes the strength of the incommensurate potential, $\phi$ is an irrational number, and $\theta\in\left[ 0 ,2\pi \right]$ is the phase angle. Similarly, we can study the evolution behavior of wavepackets in the CST-AAH model. Without loss of generality, here we discuss the case of $\sigma=1$ to study the dynamical evolution of the wavepacket by adjusting the strength of the quasiperiodic potential. Fig.~\ref{SMF2} shows the cases of $\lambda=0.1,0.5,1,1.5$, respectively. The initial position of the wavepacket is set at $j_{0}=2300$, $w=50$ and the initial momentum $p_0=-\pi/2$. The results show that wavepackets can get close to the black hole horizon for small $\lambda$. However, by increasing $\lambda$, the site where the wavepacket can reach will be further and further away from the black hole horizon ($j=0$), i.e., as $\lambda$ grows, the left region becomes even more prohibitive. To shed more light on this novel phenomenon in the CST-AAH model, we propose a ``segmentation'' method to expand our knowledge from flat to curved spacetime.

\begin{figure*}[tbhp]
	\centering
	\includegraphics[width=10cm]{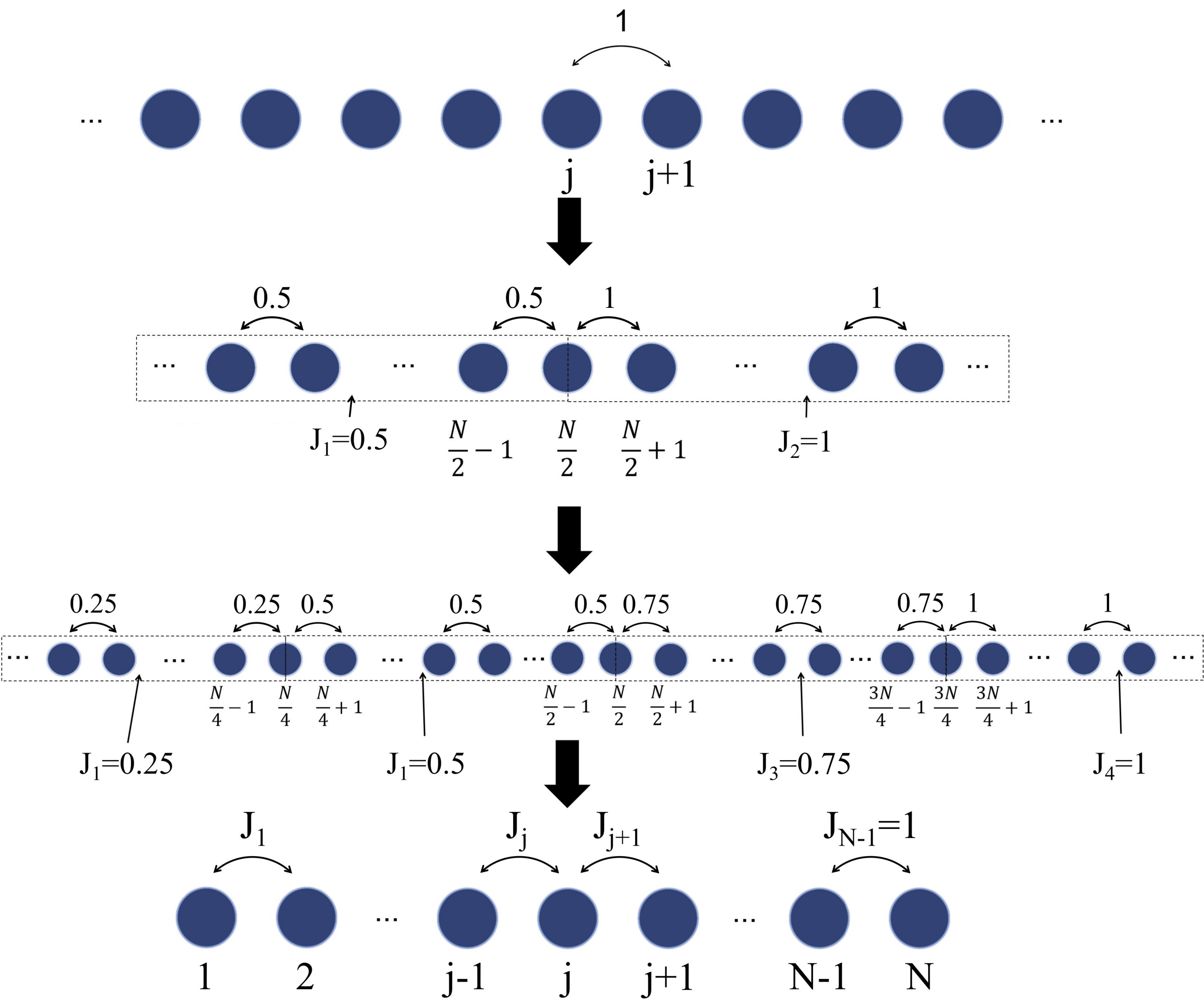}
	\caption{(color online). Schematic diagram of the segmentation method: approximating the CST-AAH model by the standard AAH model in flat spacetime. Throughout, for convenience, we choose $J=1$ as the unit of energy.} \label{S1}
\end{figure*}


\section{The segmentation method}\label{APB}
In this section, we briefly demonstrate how the segmentation method works as shown in Fig.~\ref{S1}. First, we consider a standard AAH chain with the nearest neighboring hopping strength $J$ and system size $N\rightarrow\infty$. Then we cut the chain in the middle to get two new AAH chains, with the hopping strength of the left chain being set at $0.5J$ and the right at $J$. In a system of size $N\rightarrow\infty$, the two new chains acquired by segmentation can be regarded as two individual flat-spacetime AAH models featuring different hopping strengths. Simply put, we have an AAH chain which consists of two standard flat-spacetime AAH subchains that are step-different in hopping coefficient.

In the same way as above, we cut the two subchains in the middle, respectively. That means the original AAH chain has been cut three times to become four segments, and then we set the hopping strengths of the four newly acquired subchains at $0.25J,~0.5J,~0.75J$, and $J$ from left to right. By doing so, we obtain a coupled AAH chain with three step-changes in hopping strength, which is composed of four individual flat-spacetime AAH models featuring different hopping strengths. Repeat the above segmentation $N-2$ times and we will obtain a flat-spacetime AAH model comprising $N-1$ segments with a series of hopping strengths. This reinvented AAH model can be used to theoretically analyze the CST-AAH model with $N$ sites, because the CST-AAH model with system size N also exhibits hopping strength that changes $N-1$ times. Note that, the above segmentation model is equivalent to the CST-AAH model when both the system size and the number of segmentation approach infinity. In the next section, we will take a closer look at an example of the segmentation method.

\section{Phase separation}\label{C1}
\subsection{The coupled AAH chain of two segments}
The first important discovery of the segmentation method is phase separation, where a complete AAH chain in CST is divided into two parts, with the more warped end (near a black hole) featuring localized properties while the other end (away from a black hole) maintaining extended properties. Between the localized and the extended regions exists a clear boundary, whose analytical expression, based on the segmentation method, can be obtained through simple logical deduction. Next, we will focus on the phenomenon of phase separation during the transition from flat to curved spacetime by exploring the dynamical evolution of particles in the segmentation model.

For a standard AAH model in flat spacetime, the Hamiltonian takes the form
\begin{equation}\label{AAH}
H=\sum_{j=1}^{N-1} J(\hat{c}^{\dag}_{j}\hat{c}_{j+1}+\mathrm{H.c.})+\sum_{j=1}^{N}V_{j}\hat{c}^{\dag}_{j}\hat{c}_{j},
\end{equation}
where $J$ is the hopping strength, which is equal at all sites. $\lambda$ is the strength of the quasiperiodic potential, and the rest of parameters are of the same property as in Eq.~\eqref{HAMI} (see the main text). The critical point of localized-extended phase transition for the standard AAH model is $\lambda_c=2J$, i.e., when $\lambda\le 2J$ ($\lambda\ge 2J$), all sites exhibit extended (localized) properties, simultaneously. We split it once and set the hopping strength of the left half of the chain at $0.5J$. The corresponding Hamiltonian reads
\begin{equation}
\label{Hs}
H_{s=2}=H_1+H_2,
\end{equation}
with
\begin{equation}
\label{H1}
H_{1}=\sum_{j=1}^{\frac{N}{2}}\frac{J}{2}(\hat{c}_{j}^{\dagger}\hat{c}_{j+1}+H.c.)+\sum_{j=1}^{\frac{N}{2}}\lambda \cos(2\pi \alpha j +\theta)\hat{c}_{j}^{\dagger}\hat{c}_{j},
\end{equation}
\begin{equation}
\label{H2}
H_{2}=\sum_{j=\frac{N}{2}+1}^{N-1}J(\hat{c}_{j}^{\dagger}\hat{c}_{j+1}+H.c.)+\sum_{j=\frac{N}{2}+1}^{N}\lambda \cos(2\pi \alpha j +\theta)\hat{c}_{j}^{\dagger}\hat{c}_{j},
\end{equation}
where the subscript $s=2$ indicates that the Hamiltonian consists of two parts. One can find from the Hamiltonian that the phase transition point is at $\lambda_{c1}=2J_1=2(0.5J)=J$ for a subchain satisfying $H_1$, while for $H_2$ subchain, the critical point is at $\lambda_{c2}=2(J_2)=2J$. Therefore, as the quasiperiodic potential $\lambda$ increases, $H_1$ subchain will first enter the localized phase ($J<\lambda <2J$), and then $H_2$ subchain will follow suit ($\lambda>2J$).

\begin{figure}[tbhp]
	\centering	\includegraphics[width=8.5cm]{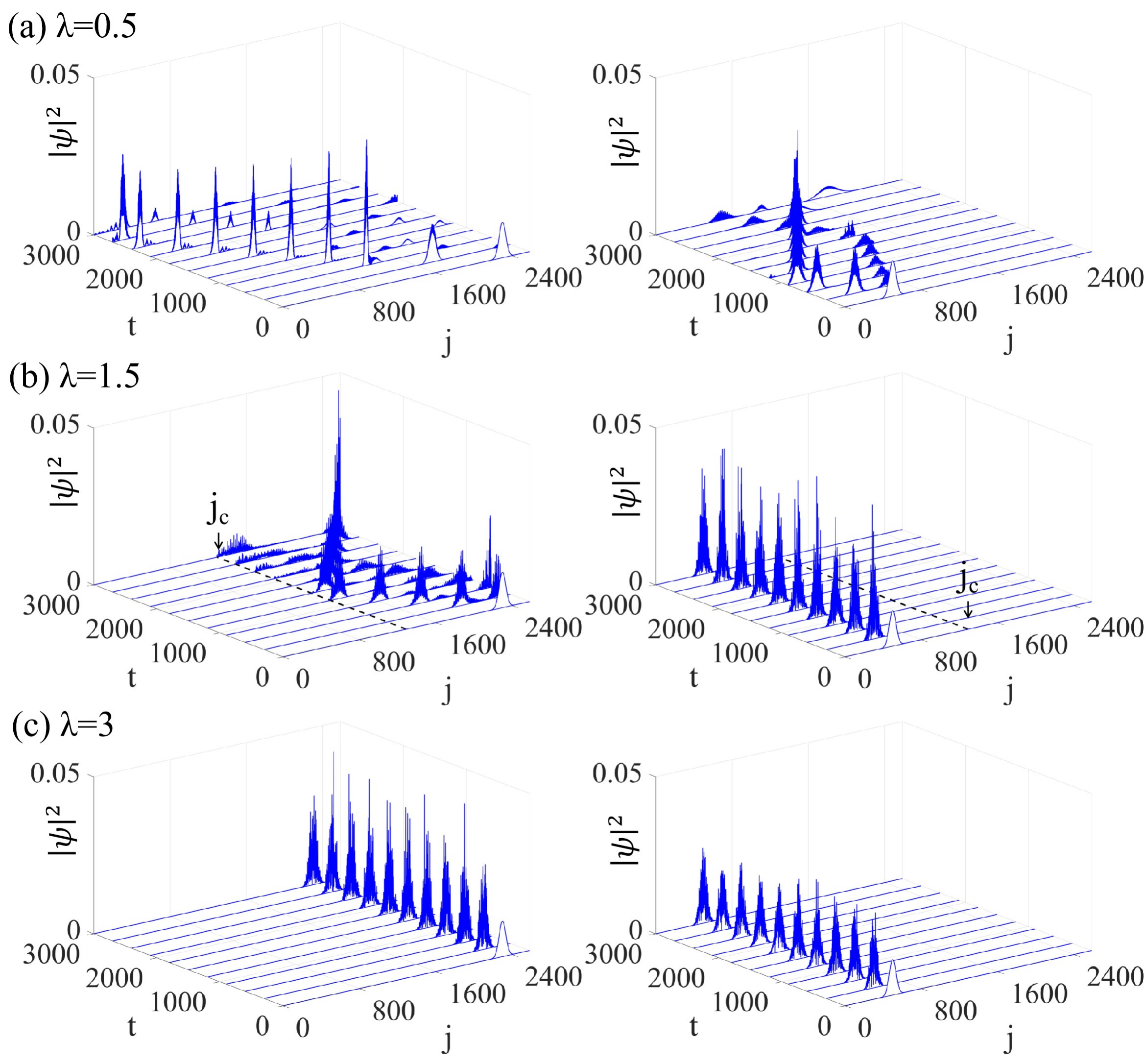}
	\caption{(color online). The wavepacket evolution for (a) $\lambda=0.5$, (b) $\lambda=1.5$ and (c) $\lambda=3$. Throughout, the initial positions of the two test wave functions are at $j_{0}=2300,~500$ and system size $N=2584$. The corresponding $j_{c}=1292$ in (b).} \label{S2}
\end{figure}

In Fig.~\ref{S2}, we show the evolution behavior of wavepackets whose initial state positions are placed in different subchains. The wavepacket whose initial state is placed in the extended subchain will never enter the localized subchain. The evolution results of wave functions again corroborate that the left half of the chain ($J_1=0.5J$) enters the localized phase first, while the right half enters the localized phase later. Remarkably, we find that when the parameters are taken between the critical points of the phase transition of the two chains ($J<\lambda <2J$), there will appear a clear boundary between the localized and extended phases [see Fig.~\ref{S2} row (b)]. This is like water and oil mixed together, which will, after being put still for a while, become spatially explicitly separated from each other. Numerical and theoretical results are consistent that when $\lambda=1.5$, the critical site of phase separation lies in the middle of the whole chain, i.e., $j_c=N/2$ [see Fig.~\ref{S2}(b)].

\subsection{The coupled AAH chain of four segments}
\begin{figure*}[tbhp]
	\centering	\includegraphics[width=17cm]{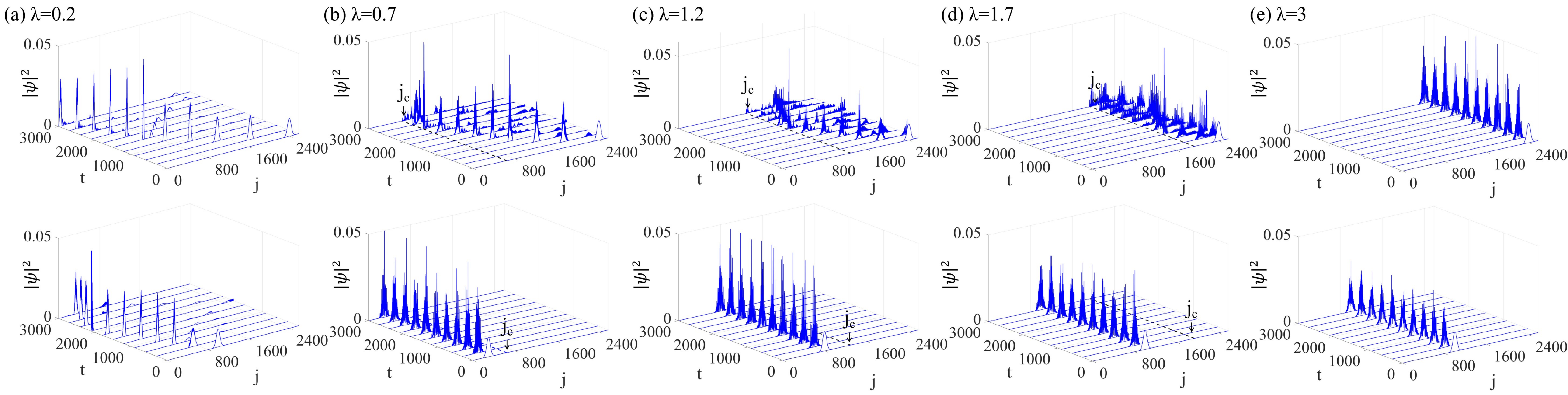}
	\caption{(color online). The wavepacket evolution for (a) $\lambda=0.2$, (b) $\lambda=0.7$, (c) $\lambda=1.2$, (d) $\lambda=1.7$ and (e) $\lambda=3$. The initial wavepacket are placed at $j_0=2400,~1000$ in column (a), $j_0=2400,~300$ in column (b), $j_0=2400,~800$ in column (c), $j_0=2400,~1000$ in column (d), and $j_0=2400,~1000$ in column (e). The system size $N=2584$. The corresponding $j_{c}=646,~1292,~1938$ in (b), (c) and(d), respectively.} \label{S33}
\end{figure*}

Now we discuss the four subchains case, where the corresponding Hamiltonian reads
\begin{equation}
\label{HHs}
H_{s=4}=H_1+H_2+H_3+H_4,
\end{equation}
with
\begin{equation}
\label{HH1}
H_{1}=\sum_{j=1}^{\frac{N}{4}}\frac{J}{4}(\hat{c}_{j}^{\dagger}\hat{c}_{j+1}+H.c.)+\sum_{j=1}^{\frac{N}{4}}\lambda \cos(2\pi \alpha j +\theta)\hat{c}_{j}^{\dagger}\hat{c}_{j},
\end{equation}
\begin{equation}
\label{HH2}
H_{2}=\sum_{j=\frac{N}{4}+1}^{\frac{N}{2}}J(\hat{c}_{j}^{\dagger}\hat{c}_{j+1}+H.c.)+\sum_{j=\frac{N}{4}+1}^{\frac{N}{2}}\lambda \cos(2\pi \alpha j +\theta)\hat{c}_{j}^{\dagger}\hat{c}_{j},
\end{equation}
\begin{equation}
\label{HH3}
H_{3}=\sum_{j=\frac{N}{2}+1}^{\frac{3N}{4}}J(\hat{c}_{j}^{\dagger}\hat{c}_{j+1}+H.c.)+\sum_{j=\frac{N}{2}+1}^{\frac{3N}{4}}\lambda \cos(2\pi \alpha j +\theta)\hat{c}_{j}^{\dagger}\hat{c}_{j},
\end{equation}
\begin{equation}
\label{HH4}
H_{4}=\sum_{j=\frac{3N}{4}+1}^{N-1}J(\hat{c}_{j}^{\dagger}\hat{c}_{j+1}+H.c.)+\sum_{j=\frac{3N}{4}+1}^{N}\lambda \cos(2\pi \alpha j +\theta)\hat{c}_{j}^{\dagger}\hat{c}_{j}.
\end{equation}
Just as the two subchains case, the phase transition points of the four new AAH subchains are $\lambda_{c1,~c2,~c3,~c4}=0.5J,~J,~1.5J,~2J$, respectively. Reusing the above analysis, we find that the system enters the localized phase in the following order: $H_1$ subchain, $H_2$ subchain, $H_3$ subchain, $H_4$ subchain. We show in Fig.~\ref{S33}, through the evolution behavior of the wavepacket over time, the phase separation phenomenon with different values of parameter $\lambda$. Other calculation parameters, like the system size etc., are marked in the figure.

Fig.~\ref{S33} shows the wavepacket dynamics of a coupled chain with four segments. The results show that when $\lambda<\lambda_{c1}$, all the subchains of the system are in the extended phase, so the initial state placed in any position can be extended throughout the entire chain [see Fig.~\ref{S33} column (a)]. As $\lambda$ increases, when $\lambda_{c1}<\lambda<\lambda_{c4}$, a part of the entire chain will enter the localized phase and the rest will be in the extended phase. Therefore, there is a clear phase separation boundary $j_{c}$ in the entire chain, and the wave function evolutions on both sides of the boundary exhibit localized and extended properties, respectively [see Fig.~\ref{S33} column (b)(c)(d)]. Furthermore, as $\lambda$ continues to increase, the entire chain becomes localized. Therefore, the wavepacket evolution exhibits localized characteristics [see Fig.~\ref{S33} column (e)].

\subsection{From the coupled AAH chain of N-1 segments to CST-AAH model}
Finally, let's turn to the case of $N-1$ segments. As shown above, under the limit of thermodynamics, subchains formed by segmentation can still be regarded as the standard AAH model in flat spacetime. Therefore, all the newly produced AAH subchains possess self-duality, which means that we can still deduce the localized-extended critical points of the subchains of the segmented AAH model from the conclusions of the standard AAH model. Since the hopping strengths difference $\Delta J=J\frac{1}{N-1}$, the critical point satisfies the expression
\begin{equation}
\lambda=2J_{j_{c}}=2J(\frac{j_{c}}{N-1}).
\end{equation}
As $\lambda$ increases, subchains satisfying $H_{j=1,2,...,N-1}$ will successively turn from the extended phase to the localized phase from left to right. If we consider the more general case where hopping difference $\Delta J$ between subchains is not a fixed value, then we need to modify the above expression as $\lambda=2J_{j_{c}}=2J(\frac{j_{c}}{N-1})^\sigma$, where $\sigma$ is the correction coefficient which controls the changing rate of hopping strength.

In thermodynamic limit, the system is divided into infinite segments. Under such circumstances, the segmentation model is equivalent to the AAH model in CST. Therefore, through similar analysis, the analytical expression of the critical site $j_c$ of the localized-extended phase separation  of the CST-AAH model with system size $N$ can be obtained as
\begin{equation}\label{ic2}
j_{c}=\left\lfloor \left(\frac{\lambda}{2J}\right)^{\frac{1}{\sigma}}(N-1) \right\rfloor,
\end{equation}
where $\left \lfloor ...\right \rfloor$ is the symbol of floor function in mathematics, which rounds down the numbers to the nearest smaller integer. At the critical site $j_c$, a clear phase boundary appears. For the fixed system size $N$, we can see from the analytical expression that $j_c$ is directly proportional to the strength of the quasiperiodic potential $\lambda$ and inversely proportional to the hopping strength $J$. Meanwhile, the CST parameter $\sigma$ determines the changing rate of the phase separation critical point $j_c$ versus $\frac{\lambda}{2J}$.

Then, let's discuss two extreme cases. We first consider the flat spacetime scenario, where the CST parameter $\sigma=0$ and the expression becomes $j_{c}=\left\lfloor \left(\frac{\lambda}{2J}\right)^{\infty}(N-1) \right\rfloor$.
As $\lambda$ grows, the system will change from an extended to a swing phase. Specifically, when $\lambda<2J$, we get $0<\frac{\lambda}{2J}<1$, and the corresponding $j_c=0$, thus the whole AAH chain is of the extended state. On the other hand, however, when $\lambda>2J$, one can get $\frac{\lambda}{2J}>1$, and then the corresponding critical site $j_c=\infty$, which leads to the localized phase in the entire AAH chain. This is in accordance with our knowledge of the standard AAH model, where no phase separation can occur.

Now let's turn to the other extreme, where the spacetime is severely warped. Without loss of generality, we take $\sigma=\infty$. From Eq.~\eqref{ic2}, we have the corresponding phase separation critical site at $j_c=N-1$, which is independent of the parameters $\lambda$ and $J$. Therefore, the extreme CST will completely freeze the hopping inside lattice systems of condensed matter to make the whole region localized.

Finally, we analyze the general case through Eq.~\eqref{ic2}. Phase separation occurs when the spacetime curvature value is finite. Taking the parameters discussed in the main text as an example, when $\sigma=1$, one can obtain the corresponding expression as $j_{c}=\left\lfloor \left(\frac{\lambda}{2J}\right)(N-1) \right\rfloor$. From the above expression, one can obtain the exact position of the site where phase separation occurs, and here the $j<j_c$($j> j_c$) part of the whole chain exhibits localized (extended) properties.

\begin{figure}[tbhp]
	\centering	\includegraphics[width=8.5cm]{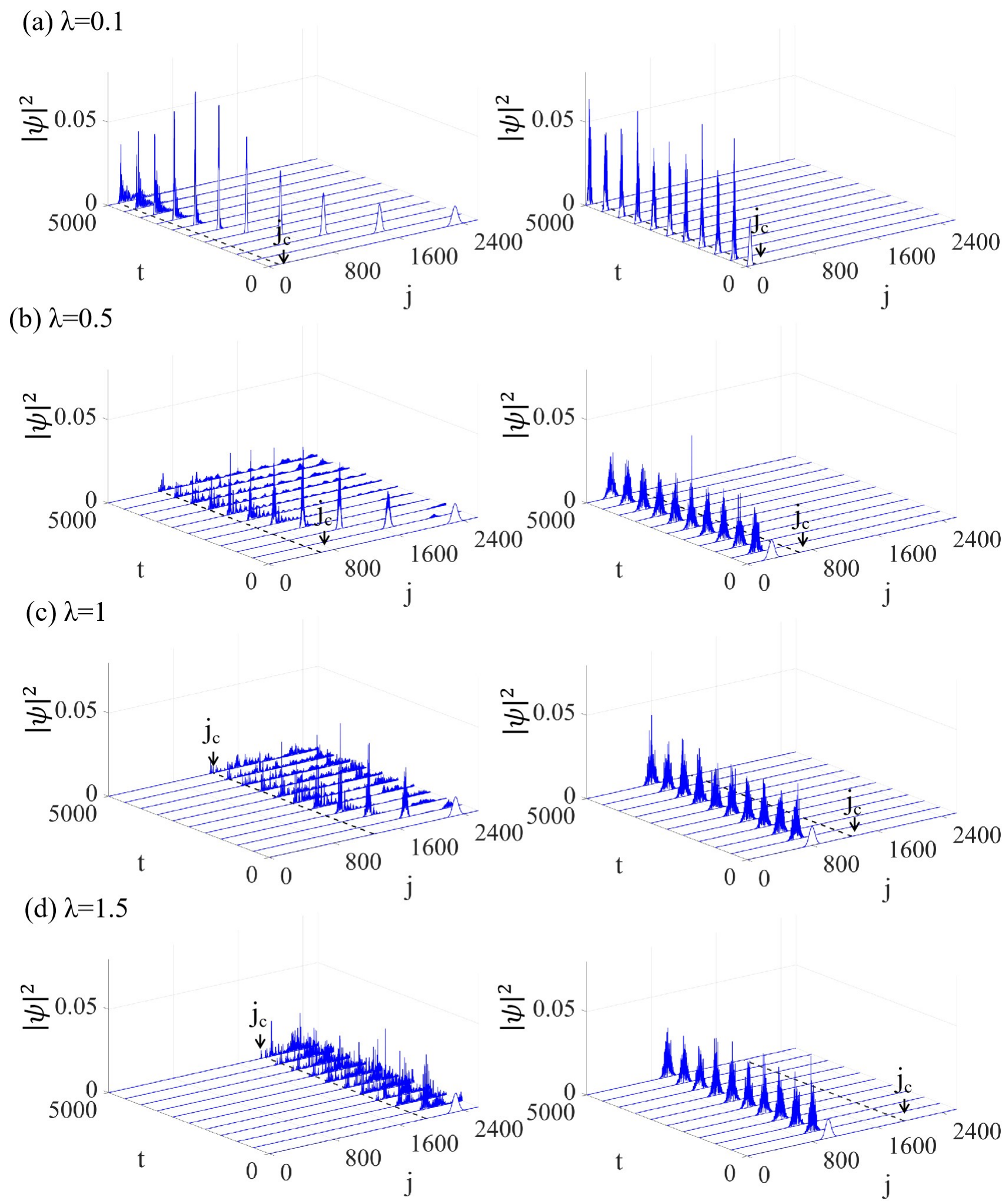}
	\caption{(color online). The wavepacket evolution of $\sigma=1$ CST-AAH model with $\lambda=0.1$ (a), $0.5$ (b), $1$ (c) and $1.5$ (d). In extended subchain, the initial positions of test wavepacket are all at $j_{0}=2300$, while for localized subchain, $j_{0}=50$, $300$, $800$ and $1000$ from the top to the bottom. For the right of row (a), we set $w=20$, while for the rest figures, $w=50$ . The system size $N=2584$ and the corresponding $j_{c}=129,~645,~1291,~1937$, respectively.}	\label{EVCST}
\end{figure}

Similarly, we can depict this characteristic of the system through the wavepacket evolution. Fig.~\ref{EVCST} shows the influence of different $\lambda$ on the position of the critical point of the phase separation. The wavepacket used for diagnosis are placed on both sides of the phase separation point, i.e., one in the extended subchain, and the other one in the localized subchain. In the above calculation, we fix the system size as $N=2584$. The results well confirm our theoretical prediction that it is CST that induces the localized-extended phase separation.

\section{Mobility edge}\label{APD}	

The occurrence of phase separation can affect the localization property of the system. In this section, we dive deeper into the mobility edge generated in the system based on the segmentation method.

\subsection{Two segments}
\begin{figure}[tbhp]
	\centering	\includegraphics[width=8.5cm]{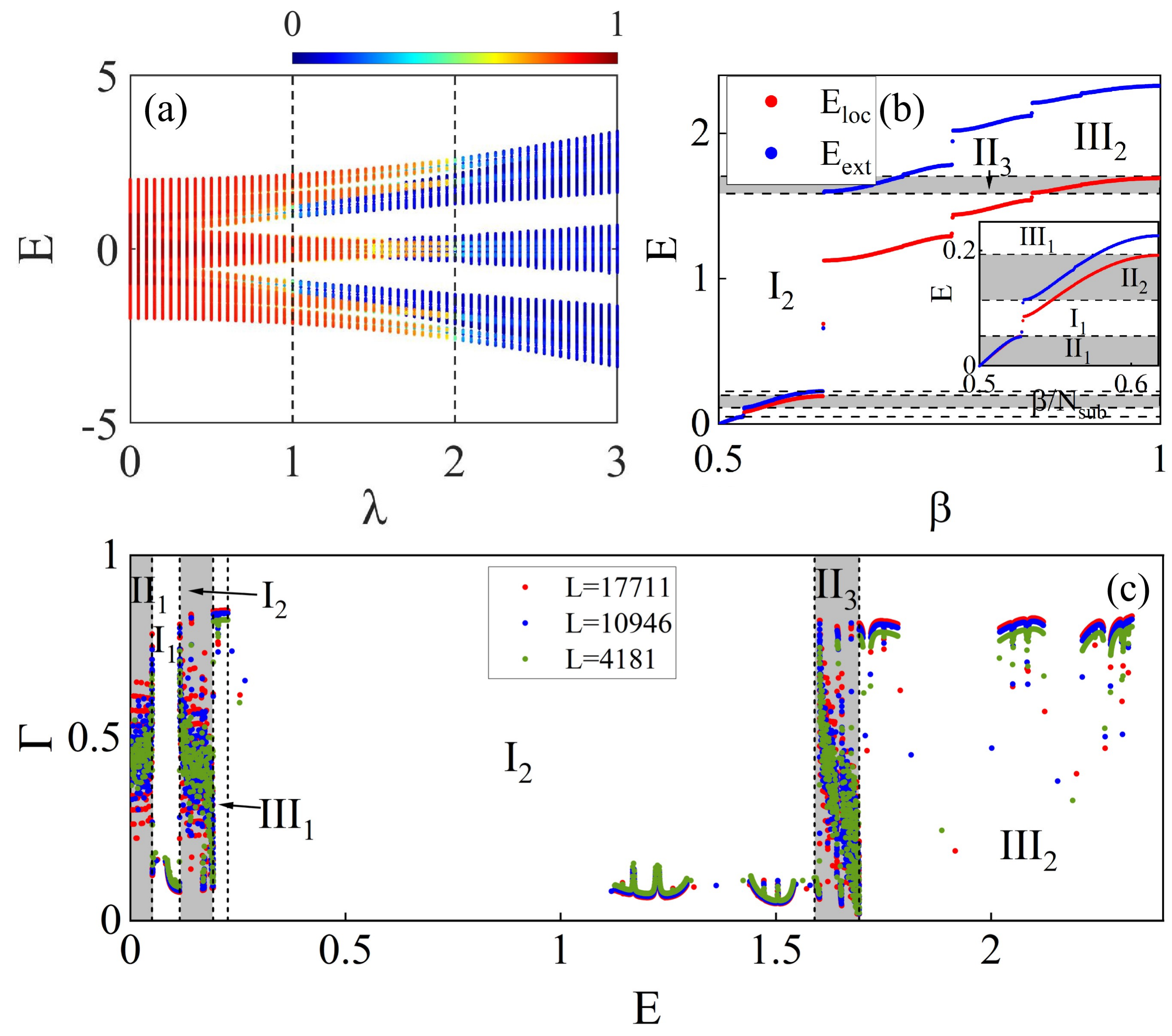}
	\caption{(color online). (a) Contour plot of eigenstate fractal dimensions corresponding to different eigenvalues as a function of $\lambda$, where the black dashed line is the transition point of the left- and right-half chain. (b) The distribution of eigenvalues versus level indices in $\lambda=1.5$. The regions where the eigenenergy of localized subsystem $E_{loc}$ and extended subsystem $E_{ext}$ overlap are colored in gray. (c) Fractal dimensions $\Gamma$ of eigenstates corresponding to different eigenvalues $E$ at different sizes $N=4181$ (green), $N=10946$ (blue) and $N=17711$ (red). The parameter $\lambda=1.5$ and system size $N=2584$ in (a) and (b). In computation, 100 times quasiperiodic averages have been performed on $\theta$.}	\label{FD1}
\end{figure}

First, we discuss the two-segment case, where fractal dimensions are calculated to characterize the localization property of the system. As shown in Fig.~\ref{FD1}(a), when the parameters $J<\lambda <2J$, the fractal dimension indicates the existence of mobility edge in the system. Previous studies have shown that there are two ways to induce mobility edge in the standard AAH model, i.e., one can either introduce long-range hopping or exert an energy-dependent quasiperiodic potential. We propose in this paper, however, a quite different approach of CST to induce mobility edge and get satisfactory results. Next, we will analyze the mobility edge through the eigenenergy distribution and scaling behavior of the system. Without loss of generality, we fix the parameter $\lambda=1.5$. In this case, the $H_1$ subchain becomes localized, while the $H_2$ subchain still remains extended. We show in Fig.~\ref{FD1}(b) the eigenenergy distribution of $E_{ext}$ and $E_{loc}$ of $H_1$ and $H_2$ subchains with respect to the level index $\beta$. The results show that there are three different phases in the system, namely, the localized phase (Region I$_{1}$ and I$_{2}$), the extended phase (Region III$_{1}$ and III$_{2}$) and the swing phase (Region II$_{1}$, II$_{2}$ and II$_{3}$). Their corresponding wave function eigenenergies show three different distributions, namely, the pure local subchain eigenvalues, the pure extended subchain eigenvalues, and the superposition of eigenvalues of the localized and extended subchains. To better understand the mobility edge, we scale the fractal dimensions of the eigenstate of the system. As shown in Fig.~\ref{FD1}(c), the fractal dimension of the localized (extended phase) decreases (increases) with the ever-growing system size, while the gray region is independent of the system size, which is the evidence of the swing phase.

\subsection{Four segments}
\begin{figure}[tbhp]
	\centering	\includegraphics[width=8.5cm]{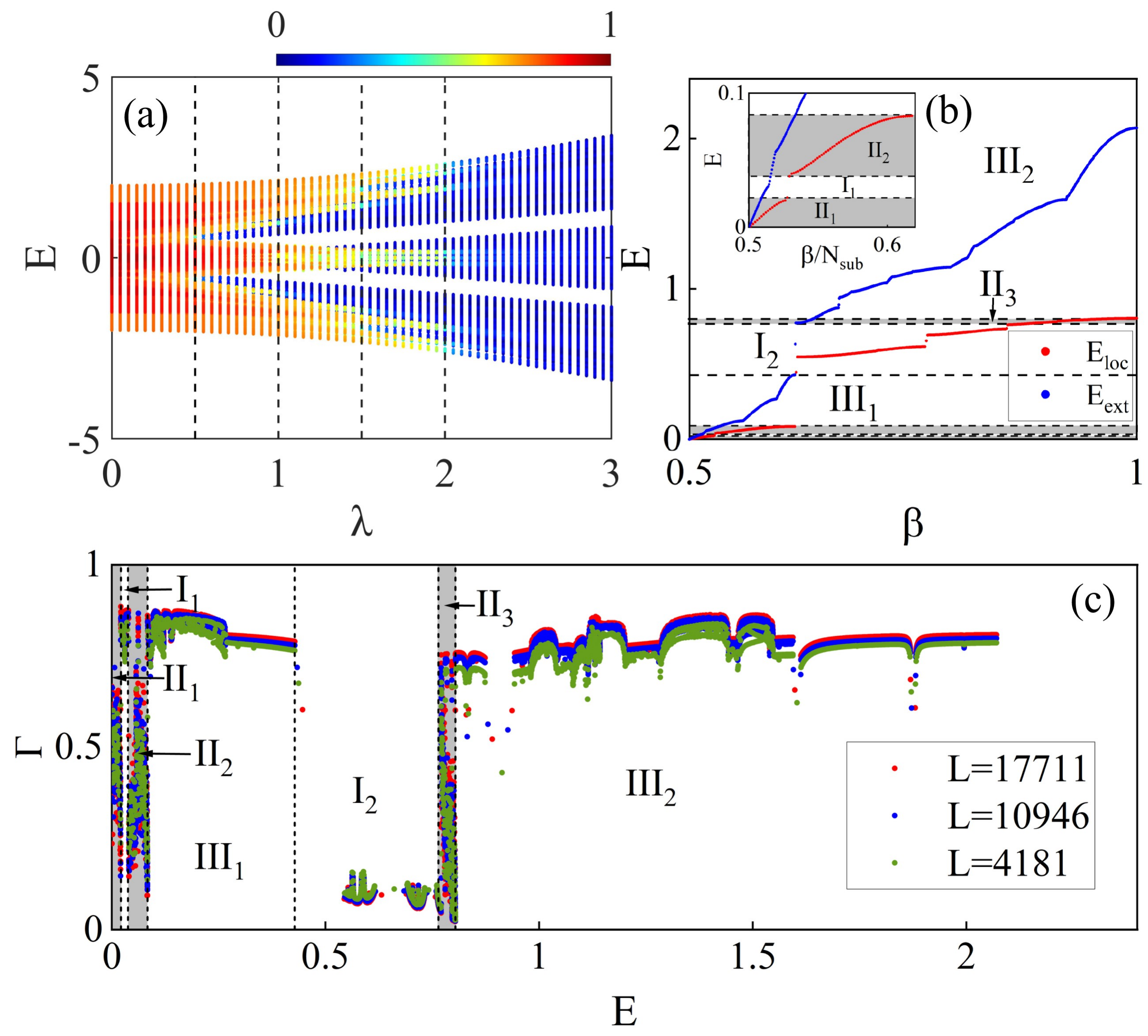}
	\caption{(color online). (a) Contour plot of eigenstate fractal dimensions corresponding to different eigenvalues as a function of $\lambda$, where the black dashed line is the transition point of $H_{1,2,3,4}$ subchain. (b) The distribution of eigenvalues versus level indices in $\lambda=0.7$. The regions where the eigenenergy of localized subsystem $E_{loc}$ and extended subsystem $E_{ext}$ overlap are colored in gray. (c) Fractal dimensions $\Gamma$ of eigenstates corresponding to different eigenvalues $E$ at different sizes $N=4181$ (green), $N=10946$ (blue) and $N=17711$ (red). The parameter $\lambda=0.7$ and system size $N=2584$ in (a) and (b). In computation, 100 times quasiperiodic averages have been performed on $\theta$.}	\label{FD2}
\end{figure}

Let's move on to the four-segment case. As has been analyzed before, since hopping strengths of the four subchains are $J_1=0.25J,~J_2=0.5J,~J_3=0.75J,~J_4=J$, respectively, they will enter the localized phase successively. Fig.~\ref{FD2}(a) shows the fractal dimension of the eigenstates, and we find that the mobility edge appears in the system when the parameter ranges $0.5J<\lambda<2J$. Since the critical points of the four subchains entering the localized phase are different, the mobility edge in the Fig.~\ref{FD2}(a) is actually composed of mobility edge of $H_{1,2,3,4}$ subchains. Then we fix $\lambda=0.7J$ and take it as an example. In this case, only $H_1$ subchain is in the localized phase while the other three subchains are all in the extended phase.

Fig.~\ref{FD2}(b) shows the eigenvalues' distribution of the localized and the extended subchain versus level indices $\beta$, and the gray area indicates where the eigenvalues overlap. The distribution of eigenvalues can clearly depict the existence of multiple energy regions in the system, which belong to three different phases, namely, the localized, the extended and the swing phases. Through the analytical expression, we calculate the critical site of phase separation at $j_c=646$. To prove the occurrence of three different phases in the system, we conduct scaling analysis again in different energy regions. The scaling behavior well reflects the differences between the regions as well as the three phases occurring in the system. This is consistent with the conclusion in the two-segment case, reaffirming that phase separation produces rich localized behaviors and mobility edge.

\begin{figure}[tbhp]
	\centering	\includegraphics[width=8.5cm]{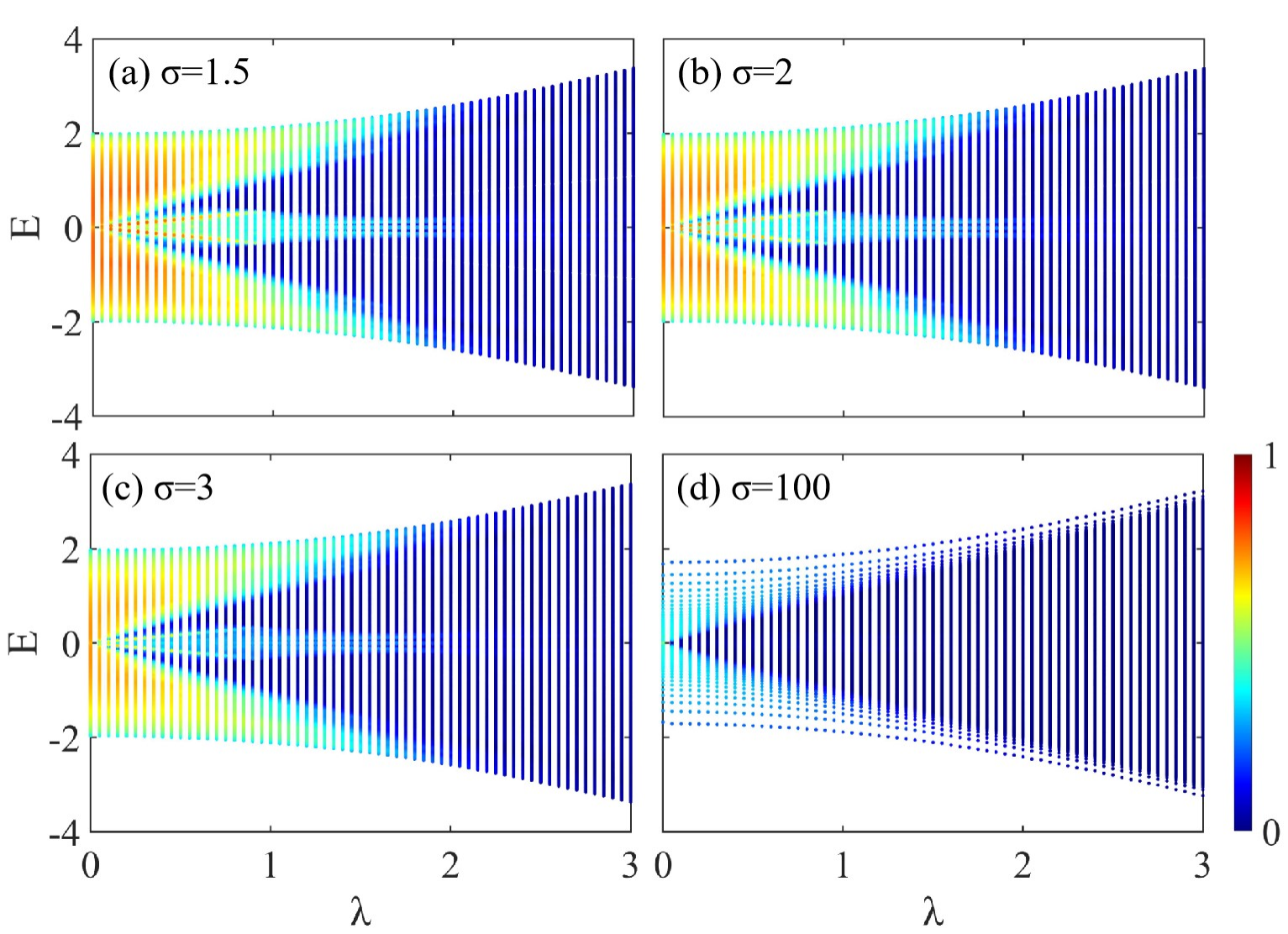}
	\caption{(color online). Fractal dimensions with different CST parameters $\sigma=1.5$ (a), $2$ (b), $3$ (c) and $100$ (d). The system size $N=2584$.}\label{S7}
\end{figure}

\subsection{N-1 segments}
Based on the above analysis, now we discuss the case of $N-1$ segments, so as to generalize the theory to CST-AAH model. One can see that in the two-segment and four-segment cases, mobility edge displayed by the fractal dimension is zigzaged because the whole chain was segmented far less than enough. However, for the $N-1$ segment case, the result of segmentation method is the same as that obtained by directly calculating the CST-AAH model, i.e., a very smooth mobility edge appears in the system. In Fig.~\ref{S7}, we illustrate how the fractal dimension of the CST-AAH model varies with the parameter $\lambda$. The results show that the mobility edge appears as the spacetime curves, and to what degree the spacetime curves has impact on the structure of the mobility edge. It is worth mentioning that since a very large CST parameter $\sigma$ will lead to a dwindling sub-extended region, the expansion of the system is inhibited. Therefore, the fractal dimension of the extended state demonstrated in the figure cannot turn red, i.e., no matter what the value of $\lambda$ is, the fractal dimension $\Gamma$ is always less than 1[see Fig.~\ref{S7}(d)].

\begin{figure}[tbhp]
	\centering	\includegraphics[width=8.5cm]{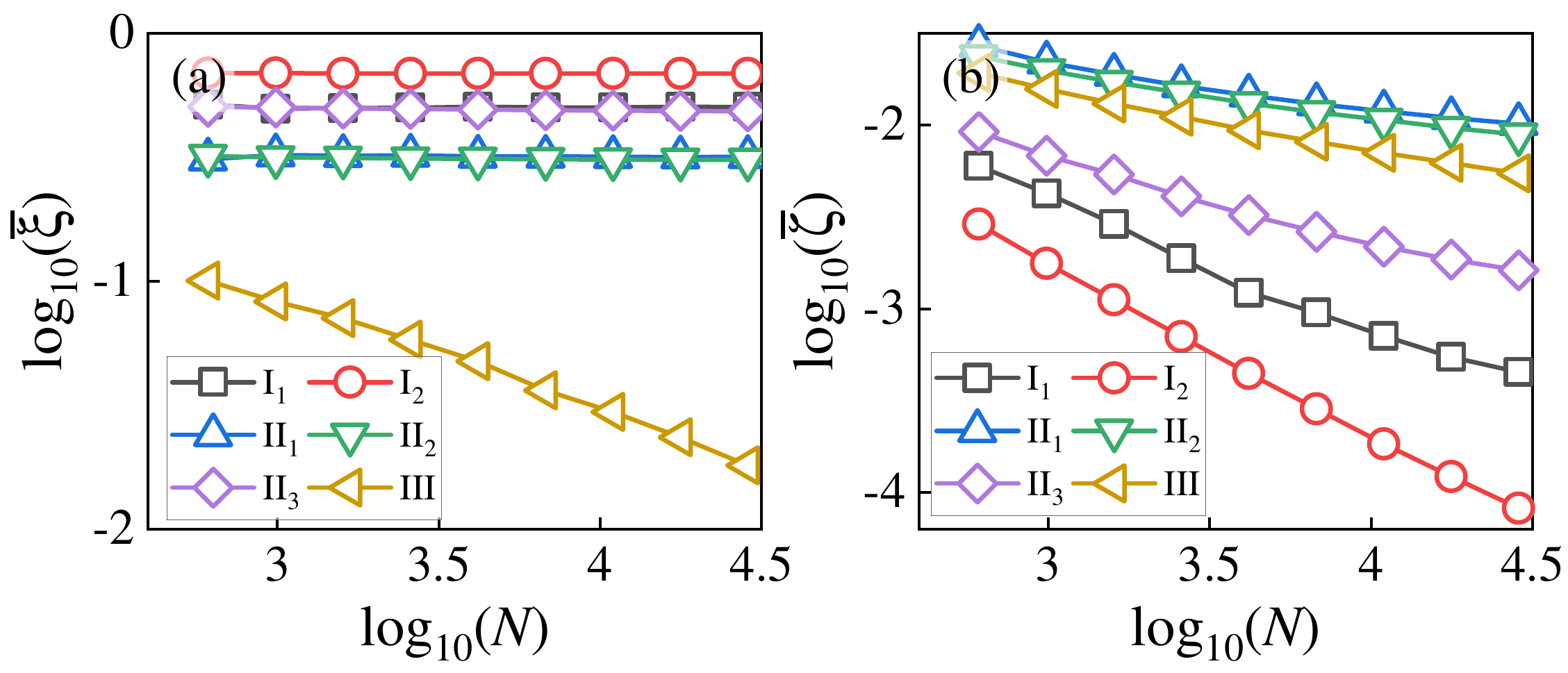}
	\caption{(color online). The scaling behavior of $\log_{10}(\bar{\xi})$ (a) and $\log_{10}(\bar{\zeta})$ (b) versus $\log_{10}(L)$ for different $\lambda$, where regions I, II, and III represent localized, swing, and sub-extended regions, respectively.}\label{SPR}
\end{figure}
\section{Scaling analysis of participation ratios}
In this section we discuss the scaling behaviour of the participation ratios in different regions for $\lambda = 1.5$, $\sigma = 1$. We define the normal participation ratio (NPR) of the $\beta$ eigenstate as
\begin{equation}
\zeta(\beta)=(N\sum_{j=1}^{N}\left | \psi_{j}(\beta)\right|^4)^{-1},
\end{equation}
For a localized (extended) state, IPR $\xi>0$ and NPR $\zeta\sim0$ (IPR $\xi\sim0$ and NPR $\zeta>0$). We define the average IPR and NPR within a region as 
\begin{equation}
\overline{\xi}=\frac{1}{\eta_{R}}\sum_{R}\xi, \ \ \ \ \
\overline{\zeta}=\frac{1}{\eta_{R}}\sum_{R}\zeta,
\end{equation}
where $\eta_{R}$ denotes the total number of eigenstates in the region $R=$I$_{1},~$I$_{2},~$II$_{1},~$II$_{2},~$II$_{3}, ~$III. The scaling behavior of the average IPR and NPR versus system size is plotted in Fig.~\ref{SPR}. In the log-log plane, the $\overline{\xi}$ of the localized and swing regions exhibit the independent of system size, whereas that of the sub-extended regions decay linearly with the increasing system size. Under certain circumstances, one cannot distinguish between the localized and swing phases. However, from the scaling behavior of the average NPR $\overline{\zeta}$, one can see that $\overline{\zeta}$ of the localized region decays linearly, whereas the $\overline{\zeta}$ of both the swing region and the sub-extended region gradually decrease to a constant with increasing system size. Comparing the IPR and NPR, one can distinguish the localized phase from the swing phase.

\begin{figure*}[tbhp]
	\centering	\includegraphics[width=16cm]{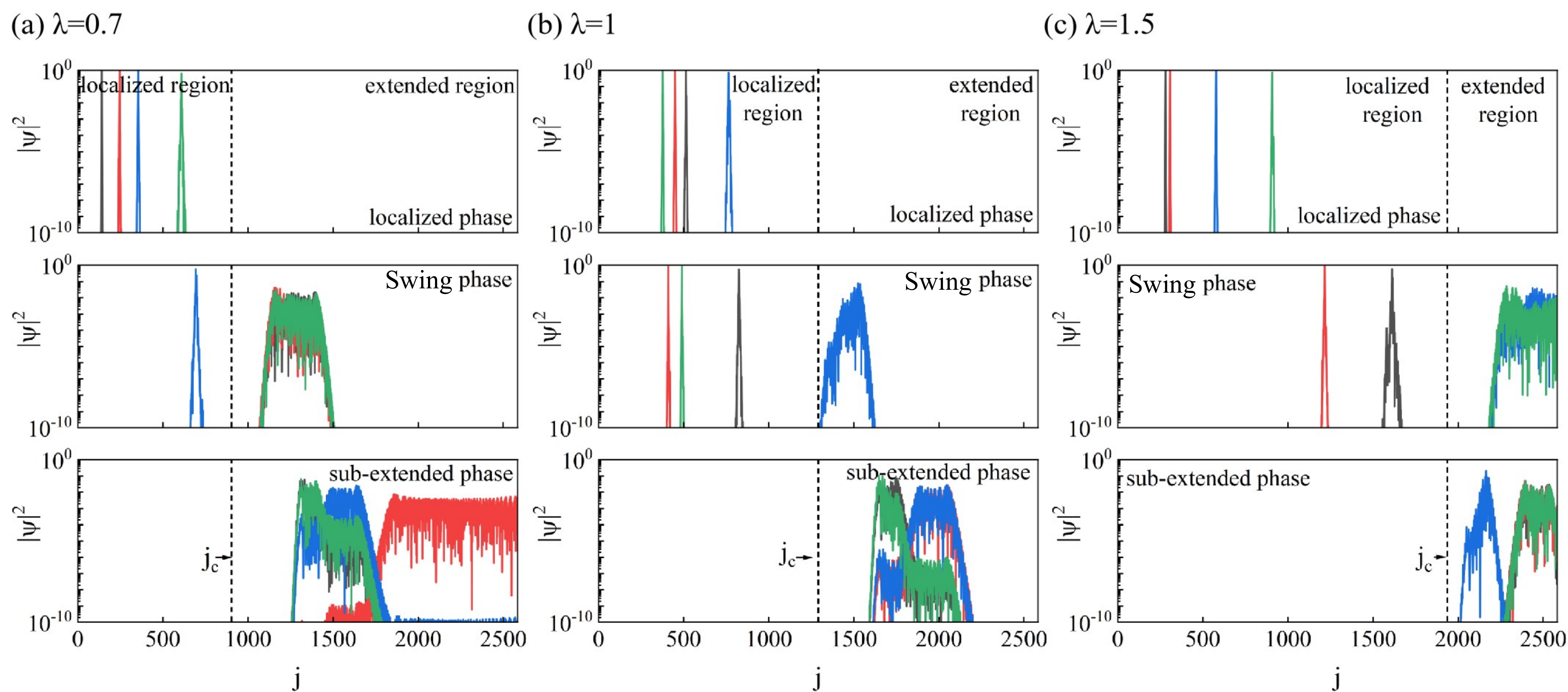}
	\caption{(color online). Density distribution of eigenstates for localized (top row), swing (middle row) and sub-extended phase (bottom row). The system size $N=2584$ and the correponding $j_c=904,~1291,~1937$ for column (a), (b) and (c), respectively.}\label{Seigen}
\end{figure*}
\section{The spatial distribution of eigenstates for different phases}
To reveal phase separation, we calculate the eigenstates of different phases versus quasiperiodic potential $\lambda$ in $\sigma=1$. As shown in Fig.~\ref{Seigen}, the localized (extended) region becomes wider (narrower) with increasing $\lambda$. The eigenstates of localized and sub-extended phases are independent of $\theta$, while that of the swing phase depend on $\theta$, which agrees well with the previous analysis.

\section{The complete phase diagram} \label{APF}
\begin{figure*}[tbhp]
\centering	\includegraphics[width=12cm]{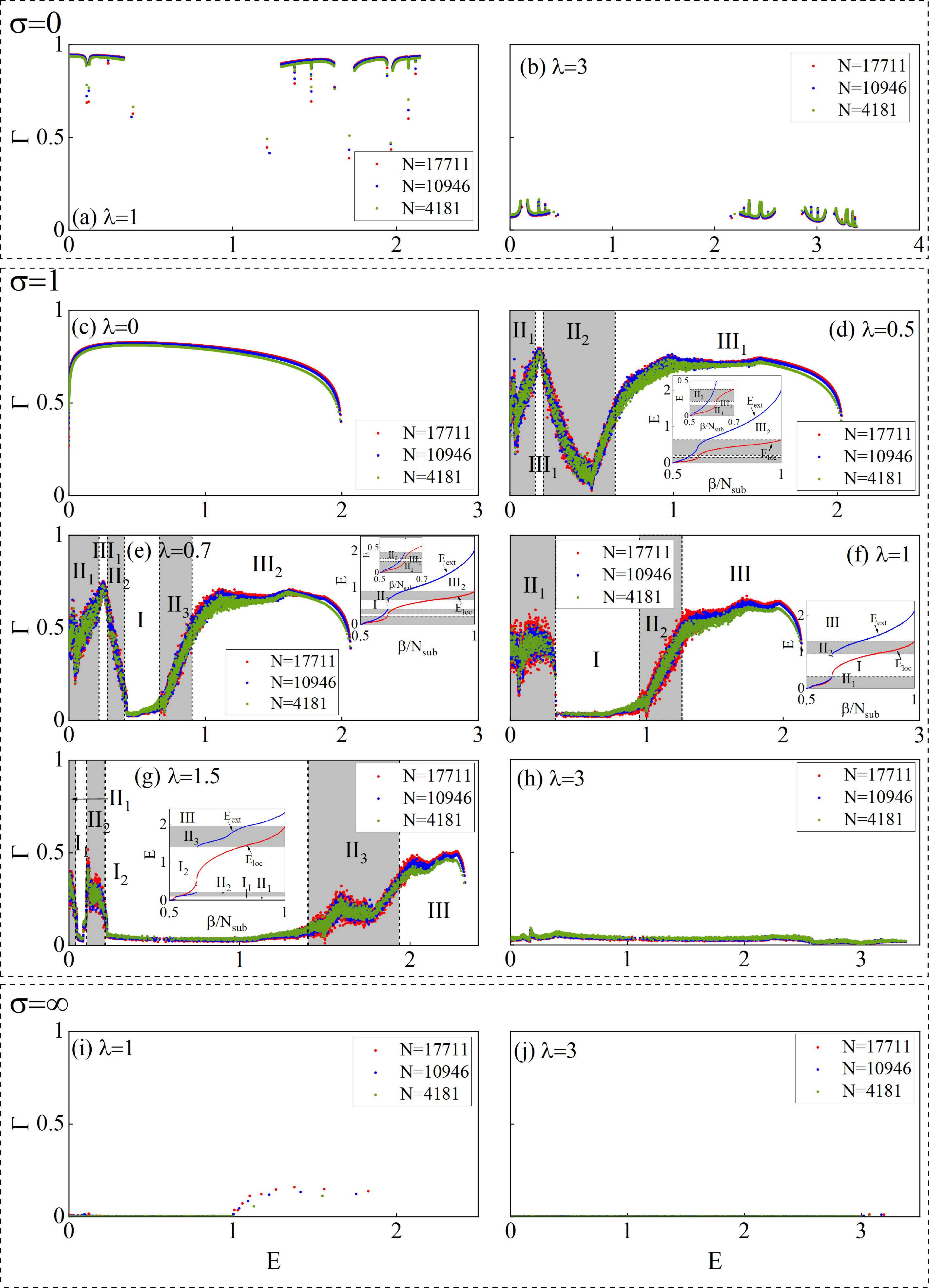}
\caption{(color online). Scaling analysis for $\sigma=0$ (traditional AAH), $\sigma=1$ (CST-AAH), $\sigma=\infty$ (the most curved case). The system size $N=4181$ (green), $N=10946$ (blue) and $N=17711$ (red). The quasiperiodic potential strength $\lambda=0$ (a), $\lambda=1$ (b), $\lambda=0$ (c), $\lambda=0.5$ (d), $\lambda=0.7$ (e), $\lambda=1$ (f), $\lambda=1.5$ (g), $\lambda=3$ (h), $\lambda=1$ (i), and $\lambda=3$ (j). The insets show the eigenvalues $E_{loc}$ ($E_{ext}$) of the localized (extended) subsystem versus different energy indices $\beta$. In computation, 100 times quasiperiodic averages have been performed on $\theta$.}\label{S8}
\end{figure*}

In the end, we provide a complete phase diagram of AAH model in CST by fractal dimensions under different parameters. In Fig.~\ref{S8}, by showing scaling behaviors of the fractal dimension at different energy $E$, we list and summarize all possible phases of the CST-AAH model in Fig.~4 in the main text. Different dashed boxes in Fig.~\ref{S8} correspond to the results of $\sigma=0$ (flat spacetime), $\sigma=1$ (CST) and $\sigma=\infty$ (CST to the extreme), respectively. One can see that although the CST-AAH model still experiences a change from the extended to the localized phases as  the quasiperiodic potential $\lambda$ increases, the process of phase transition differs from AAH model in flat spacetime. To be specific, during phase transition of the CST-AAH model there appears four transitory stages, i.e., the extended phase steps into the swing phase first, and then gradually enters the localized phase.

\begin{figure*}[tbhp]
	\centering	\includegraphics[width=12cm]{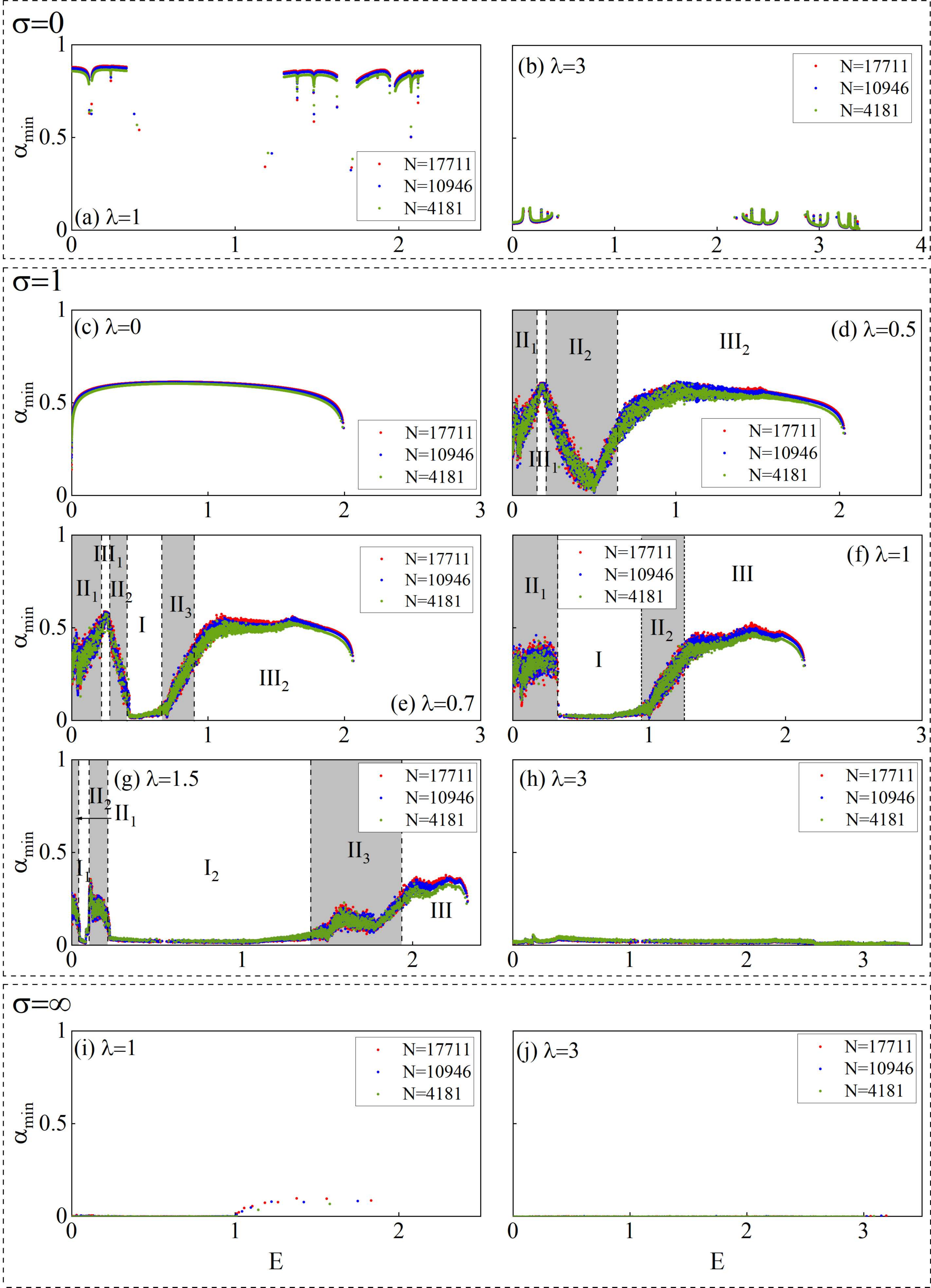}
	\caption{(color online). The minimum of the scaling index $\alpha_{min}$ for $\sigma=0$ (traditional AAH), $\sigma=1$ (CST-AAH), $\sigma=\infty$ (the most curved case). The system size $N=4181$ (green), $N=10946$ (blue) and $N=17711$ (red). The quasiperiodic potential strength $\lambda=0$ (a), $\lambda=1$ (b), $\lambda=0$ (c), $\lambda=0.5$ (d), $\lambda=0.7$ (e),  $\lambda=1$ (f), $\lambda=1.5$ (g), $\lambda=3$ (h), $\lambda=1$ (i), and $\lambda=3$ (j). In computation, 100 times quasiperiodic averages have been performed on $\theta$.}\label{SI}
	\end{figure*}

Another way to distinguish among the three different phases and confirm the mobility edge is to examine the wave function itself. This method is called multifractal analysis and is often used to study the localized behavior of AAH models ~\cite{HGrussbac1995}. Through the scaling index in the multifractal analysis, we again test the correctness of the above results. The concrete calculation results are shown in Fig.~\ref{SI}. The fractal dimension and scaling index consistently prove the existence of mobility edge in CST-AAH model.

%
\end{document}